\documentclass[pdflatex,sn-nature]{sn-jnl}%

\usepackage{graphicx}%
\usepackage{multirow}%
\usepackage{amsmath,amssymb,amsfonts}%
\usepackage{amsthm}%
\usepackage{mathrsfs}%
\usepackage[title]{appendix}%
\usepackage{xcolor}%
\usepackage{textcomp}%
\usepackage{manyfoot}%
\usepackage{booktabs}%
\usepackage{algorithm}%
\usepackage{algorithmicx}%
\usepackage{algpseudocode}%
\usepackage{listings}%
\usepackage{gensymb}%
\usepackage{siunitx}

\usepackage{tabularx}
\usepackage{array}
\usepackage{makecell}
\usepackage[section]{placeins}
\newcolumntype{L}{>{\raggedright\arraybackslash}X}
\newcolumntype{C}[1]{>{\centering\arraybackslash}m{#1}}

\begin{document}

\title[Multi-scale reconstruction of single-ion damage tracks in diamond via nitrogen-vacancy centers]{Multi-scale reconstruction of single-ion damage tracks in diamond via nitrogen-vacancy centers}

\author*[1]{\fnm{Daniel G.} \sur{Ang}}\email{dga@umd.edu}
\equalcont{These authors contributed equally to this work.}
\author*[1,2]{\fnm{Jiashen} \sur{Tang}}\email{jstang@umd.edu}
\equalcont{These authors contributed equally to this work.}
\author[1,3]{\fnm{Maximilian} \sur{Shen}}
\author[4]{\fnm{Michael} \sur{Titze}\,$^{\ddagger}$\footnotetext{$^{\ddagger}$Present address: Advanced Instrumentation for Nano-Analytics, Luxembourg Institute of Science and Technology, Belvaux L-4422, Luxembourg.}}
\author[1,3]{\fnm{Gavishta} \sur{Liyanage}}
\author[5]{\fnm{Jordan} \sur{Chapman}}
\author[1,2]{\fnm{Chinmay} \sur{Bharathulwar}}
\author[1,2]{\fnm{Andrew T.} \sur{Gilpin}}
\author[6, 9]{\fnm{Tanguy} \sur{Terlier}}
\author[4]{\fnm{Edward S.} \sur{Bielejec}}
\author[5,7,8]{\fnm{Vsevolod} \sur{Ivanov}}
\author*[1,2,3]{\fnm{Ronald L.} \sur{Walsworth}}\email{walsworth@umd.edu}

\affil[1]{\orgname{Quantum Technology Center, University of Maryland}, \orgaddress{\city{College Park}, \postcode{20742}, \state{Maryland}, \country{USA}}}
\affil[2]{\orgdiv{Department of Physics}, \orgname{University of Maryland}, \orgaddress{\city{College Park}, \postcode{20742}, \state{Maryland}, \country{USA}}}
\affil[3]{\orgdiv{Department of Electrical Engineering and Computer Science}, \orgname{University of Maryland}, \orgaddress{\city{College Park}, \postcode{20742}, \state{Maryland}, \country{USA}}}
\affil[4]{\orgname{Sandia National Laboratories}, \orgaddress{\city{Albuquerque}, \postcode{87123}, \state{New Mexico}, \country{USA}}}
\affil[5]{\orgname{Virginia Tech National Security Institute}, \orgaddress{\city{Blacksburg}, \postcode{24060}, \state{Virginia}, \country{USA}}}
\affil[6]{\orgname{SIMS Laboratory, Shared Equipment Authority, Rice University}, \orgaddress{\city{Houston}, \postcode{77005}, \state{Texas}, \country{USA}}}
\affil[7]{\orgdiv{Department of Physics}, \orgname{Virginia Tech}, \orgaddress{\city{Blacksburg}, \postcode{24061}, \state{Virginia}, \country{USA}}}
\affil[8]{\orgname{Virginia Tech Center for Quantum Information Science and Engineering}, \orgaddress{\city{Blacksburg}, \postcode{24061}, \state{Virginia}, \country{USA}}}
\affil[9]{\orgdiv{Department of Chemical and Biomolecular Engineering}, \orgname{Rice University}, \orgaddress{\city{Houston}, \postcode{77005}, \state{Texas}, \country{USA}}}

\abstract{Understanding particle-induced damage tracks in solid-state materials underpins emerging applications in rare-event detection and quantum defect engineering. Resolving these tracks requires multi-scale readout, from event localization at the millimeter scale to track-morphology reconstruction at the nanoscale. Nitrogen-vacancy (NV) centers in diamond provide such a platform, combining optical localization with quantum sensing of track morphology. Here, we implant sub-MeV carbon ions into nitrogen-rich diamond and detect individual recoil events via spatially localized NV formation. We develop a simulation framework that explains the observed NV yield and predicts that directional information is retained in the NV distribution after annealing. Machine learning further recovers much of the information lost to defect diffusion and limited NV yield, improving head-tail classification to a level comparable to pre-annealed vacancy tracks. Measurements of NV spin coherence indicate compatibility with nanoscale track reconstruction via NV strain mapping and magnetic gradient-based techniques. These results identify promising pathways toward NV-diamond directional detectors for rare events, while the track-modeling framework has broader implications for paleodetection and quantum material synthesis.}


\maketitle

\begin{figure}[htbp]
    \centering
    \includegraphics[width=\textwidth]{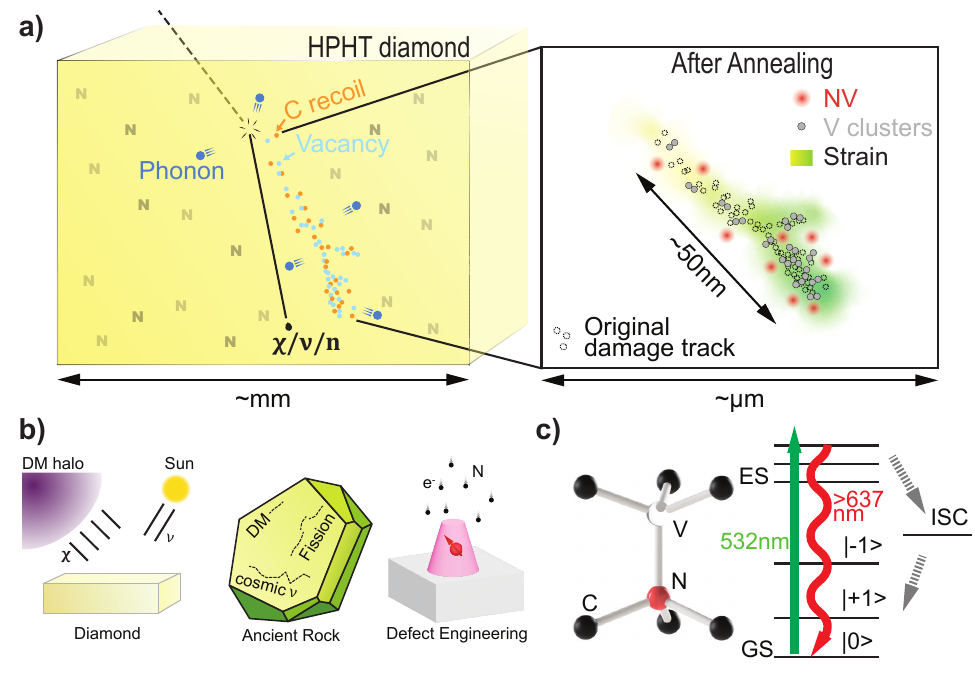}
    \caption{Concept of NV-based directional rare-event detection in diamond. a) Schematic of damage track formation and multi-scale readout in nitrogen-rich HPHT diamond. An incoming particle induces a carbon nuclear recoil, generating a cascade of vacancies and interstitials. The recoil also produces phonons that enable identification of candidate interaction events at the millimeter scale (single diamond chip). During annealing, vacancies diffuse and either recombine with interstitials, form vacancy clusters, or combine with substitutional nitrogen to create optically-active NV centers near the original track. The resulting NV photoluminescence (PL) enables localization at the micrometer scale, while the spatial distribution of NV centers and residual defect-induced strain enables nanoscale reconstruction of track morphology and inference of the incoming particle direction. b) Directional information enables discrimination between dark matter signals and solar neutrino backgrounds. More broadly, studies of damage track formation and evolution are relevant for paleodetector-based searches in ancient minerals and for the formation of quantum defects via implantation. c) Crystal structure and energy levels of the NV center. The NV center is a spin-1 defect that can be optically initialized with green light ($\sim$532\,nm) and read out via spin-dependent PL ($>$637 nm).} 
    \label{fig:intro}
\end{figure}

The study of damage tracks formed by particle-lattice interactions in solid-state materials has played a fundamental role in diverse fields such as geoscience \cite{tagami2005}, particle detection \cite{fleischer1965solid}, radiation damage in structural materials for nuclear and fusion energy \cite{nordlund2018}, carbon ion radiotherapy \cite{Karger_2018}, and swift heavy ion irradiation \cite{lang2020fundamental,amekura_latent_2024,lake_direct_2021,liu_optical_2025}. Recently, interest in damage tracks has expanded into two frontier areas of fundamental particle physics and quantum technology. Tracks preserved in ancient minerals \cite{baum_mineral_2023,Drukier_DMminerals_2019} or ultrapure crystals \cite{rajendran_method_2017,araujoNuclearRecoilDetection2025} have been proposed as probes of rare events such as neutrino and dark matter (DM) interactions, while controlled track formation has become integral to engineering the properties of color centers in quantum materials \cite{favarodeoliveiraTailoringSpinDefects2017,luhmann2019coulomb, Delegan_2023,kim2025scalable}. These emerging applications require the ability to model and characterize particle-induced damage tracks, including their morphology, defect generation, and impact on the surrounding lattice environment. The negatively charged nitrogen-vacancy center in diamond (hereafter referred to as the NV center) provides a quantum sensor platform that addresses this need directly.
NV centers can be optically resolved at the single-defect level \cite{dohertyNitrogenvacancyColourCentre2013}, while their spin properties enable high sensitivity to local lattice strain \cite{kehayias_crystalstress_2019,marshall_high-precision_2022,Ang2025} and nanoscale spatial mapping \cite{arai_fourier_2015,zhang_selective_2017,amawiThreedimensionalMagneticResonance2024}. Advances in NV quantum sensing have established protocols for high-throughput, high-sensitivity imaging across a broad range of physical quantities~\cite{levine_principles_2019,BarrySensOpt2020}.
These capabilities enable NV centers to probe nearby damage tracks (Fig. \ref{fig:intro}a) by providing access to their local structure, associated lattice perturbations, and importantly, unlocking the potential to reconstruct the direction of the incoming particle~\cite{rajendran_method_2017}.  
Such directional sensitivity is a central capability for a proposed diamond-based directional dark matter detector~\cite{marshall_directional_2021,ebadi_directional_2022}, as it will allow discrimination of the DM signal from solar neutrino backgrounds (Fig. \ref{fig:intro}b), which are beginning to limit conventional Weakly Interacting Massive Particle (WIMP) searches~\cite{ohareNewDefinitionNeutrino2021,XENONsolarneutrino2024}. A solid-state directional detector is particularly attractive as it will require much less detector volume compared to traditional gas-based directional detectors~\cite{vahsenDirectionalRecoilDetection2021} while building directly on the NV sensing capabilities described above.
Diamond-based rare event detectors can also probe sub-GeV light dark matter in a non-directional mode, capitalizing on diamond's favorable properties for conventional signal readout~\cite{umemoto2023scintillation,kurinsky_diamond_2019}. Finally, directional diamond-based neutrino detection capabilities can also improve tests of the Standard Model at high-flux, energetic neutrino sources~\cite{strigaricevnsdirectional2020}.

Realizing such a detector requires multi-scale readout of individual damage events, spanning millimeter, micrometer, and nanometer length scales~\cite{Ang2024}. A DM or neutrino interaction induces a nuclear recoil \cite{rajendran_method_2017}, generating a nanoscale cascade of lattice defects -- vacancies and interstitials -- whose spatial distribution encodes information about the incoming particle direction. Candidate events are first identified at the millimeter scale using conventional phonon, charge, and/or photon readout \cite{kurinsky_diamond_2019,kim2025athermal,umemoto2023scintillation}, followed by micrometer-scale localization via wide-field NV strain imaging using pre-existing NV ensembles or optical detection of newly formed NV centers after annealing~\cite{marshall_directional_2021,marshall_high-precision_2022,Ang2025}. At the nanoscale, the morphology of the damage track may be probed using techniques such as high-resolution X-ray strain imaging \cite{marshall_xray_2021, wierbik2026anisotropic} or Fourier-gradient imaging of individual NV centers \cite{zhang_selective_2017} to reconstruct the original track direction. These approaches together constitute a multi-scale readout pipeline, but the optimal sensing modality at each length scale remains an open question. Determining which techniques are viable requires better quantitative understanding of damage track properties, including defect formation, annealing evolution, and the resulting optical or strain signatures.

Here, we present an experimental and computational study of single-ion carbon implantation in synthetic high-pressure, high-temperature (HPHT) diamond to inform multi-scale track reconstruction. HPHT diamonds have a high concentration of nitrogen impurities and a low density of pre-existing NV centers, enabling implantation-generated vacancy defects to form NV centers during annealing in proximity to the original damage track. We experimentally demonstrate detection of individual implanted sub-MeV ions via spatially localized photoluminescence (PL) at the micrometer scale. We then develop a computational model of track formation and NV creation that matches the measured NV yield and predicts the spatial distribution of NV centers, including the directional information retained at the nanoscale. Measurements of NV spin coherence further indicate compatibility with nanoscale reconstruction of damage tracks via magnetic gradient-based quantum sensing. Our study helps constrain possible pathways toward diamond-based directional rare-event detectors. More broadly, the framework could be extended to study track formation and survivability over geological timescales relevant to mineral-based dark matter detection (Fig. \ref{fig:intro}b)~\cite{Drukier_DMminerals_2019,baum_mineral_2023}, and to guide implantation strategies for quantum device fabrication.

\section{Results}
\subsection{Ion implantation and optical characterization of nitrogen-vacancy color centers}

\begin{figure}[htbp]
    \centering
    \includegraphics[width=1\textwidth]{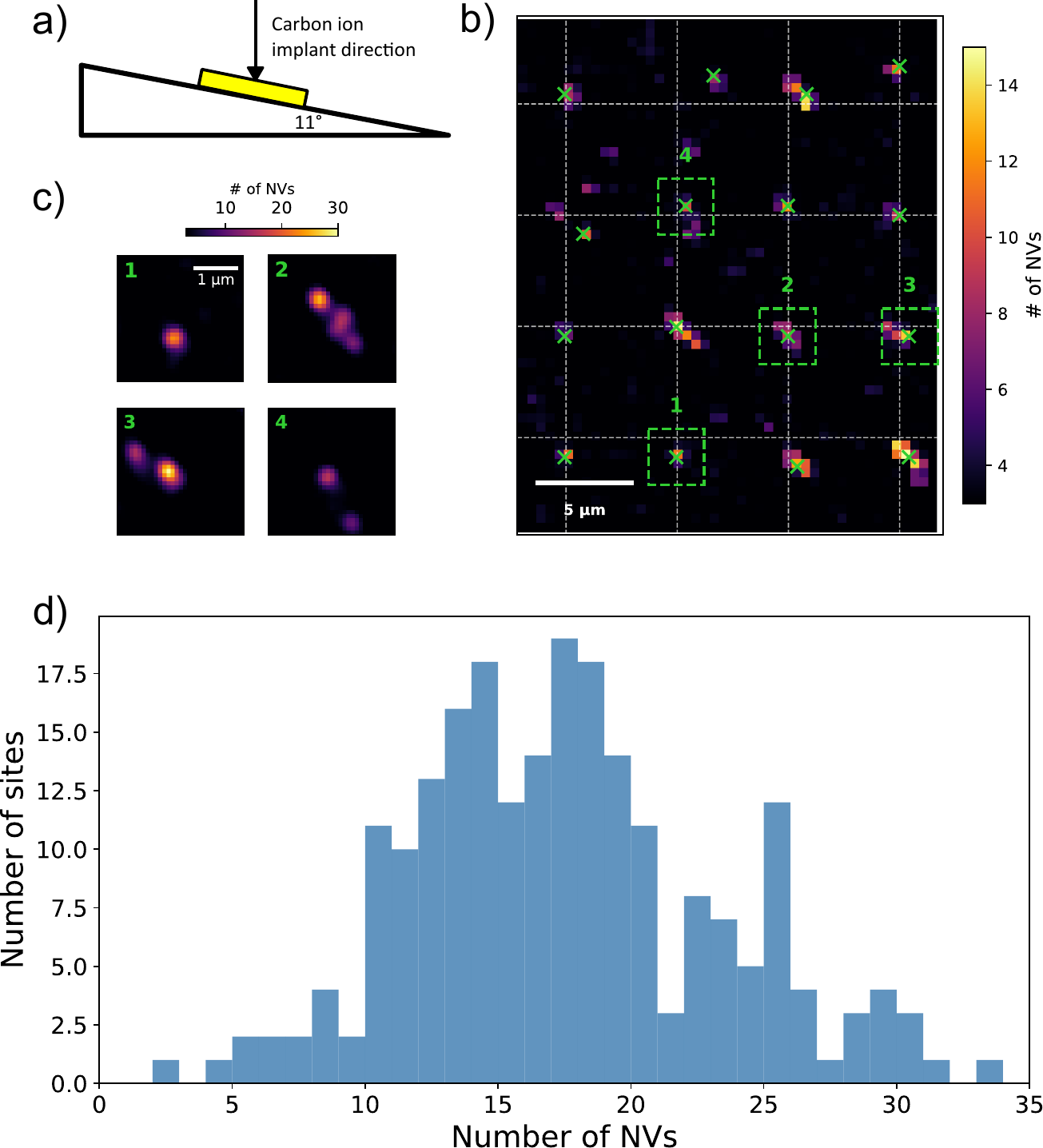}
    \caption{Ion implantation and optical characterization of NV formation. a) Schematic of ion implantation geometry. The diamond surface is tilted by ${\sim}11\degree$ relative to the incident carbon ion beam to suppress channeling effects. b) Wide-field confocal NV PL map acquired at coarse spatial resolution ($\sim$500\,nm). Bright PL spots correspond to NV ensembles formed at individual implantation sites, arranged in a regular grid consistent with the implantation pitch. Candidate sites selected for further high-resolution 3D confocal scans are marked with green crosses. The NV number is estimated from the PL intensity relative to calibrated single NV. Background NV counts in the HPHT sample are $\lesssim2$. c) Representative high-resolution confocal scans of four implantation sites (fixed Z slice). Spatially separated PL spots are observed and are attributed to multi-ion implantation events, consistent with statistical analysis. d) Distribution of the number of NV centers per implanted ion ($N_{\rm NV}$), extracted from integrated PL intensity (see main text). Data are collected from 222 sites, yielding a mean NV count of ($17.5\pm0.38$), where the uncertainty denotes one standard error of the mean; the standard deviation is 5.7.} 
    \label{fig:confocalscanres}
\end{figure}

Carbon ions with an energy of 800\,keV were implanted into bulk HPHT diamond at the Sandia National Laboratories microbeam facility. This energy places the resulting damage track $\sim$600\,nm deep into the surface, suppressing surface-related noise and defect outdiffusion~\cite{rackeVacancyDiffusionNitrogenvacancy2021} to produce a sufficient number of NV centers for optical detection while being near to the $<$500 keV energy range relevant to nuclear recoils induced by WIMP dark matter and neutrinos. Implantation was performed in square grids with the surface tilted by 11 degrees to suppress channeling effects (Fig. \ref{fig:confocalscanres}a). Following implantation, high temperature annealing was performed to induce the migration of vacancies into nearby nitrogen atoms, resulting in NV center formation near the individual implantation sites. The high concentration of substitutional nitrogen in the HPHT diamond, measured to be $\sim$253\,ppm (see Supplementary Information), provides an electron-donor-rich environment that favors the negatively charged NV$^-$ state \cite{deakFormationNVCenters2014}; in similarly nitrogen-dense diamond, NV$^-$ has been reported to dominate over NV$^0$ by approximately two orders of magnitude \cite{manson2018}. A coarse PL scan using a confocal setup was then performed, revealing arrays of bright PL spots with spacing matching the implantation pitch (6 or 10 $\mu$m), with spatial variations reflecting the beam size uncertainty (Fig. \ref{fig:confocalscanres}b; see Methods). These results demonstrate that individual carbon ion impacts generate spatially localized ensembles of NV centers that can be optically resolved.

To quantify the NV yield per ion, candidate implantation sites were identified using an automated grid-matching procedure, which also rejects backgrounds from surface polishing damage (see Supplementary Information). High-resolution three-dimensional confocal scans were then acquired for each identified site. For sites that received multiple ions, we observed spatially separated PL spots (Fig. \ref{fig:confocalscanres}c). The occurrence of such multi-spot events is consistent with a zero-truncated Poisson distribution, confirming that each PL spot within a multi-spot grid point corresponds to a single-ion impact (see Supplementary Information). The fitted Poisson rate of $\lambda = 0.99$ indicates that approximately one ion was delivered per nominal grid point, consistent with the experimental beam-current settings. The number of NV centers per implanted ion, $N_{\rm NV}$, was then determined by comparing the integrated PL intensity of each spot to that of calibrated single NV centers under identical imaging conditions. The resulting distribution of $N_{\rm NV}$ across 222 implantation sites is shown in Fig. \ref{fig:confocalscanres}d, with a mean value of $17.5\pm0.38$ (standard error of the mean; with site-to-site standard deviation $=5.7$).

\subsection{Modeling NV formation after implantation and annealing}

\begin{figure}[htbp]
    \centering
    \includegraphics[width=0.9\textwidth]{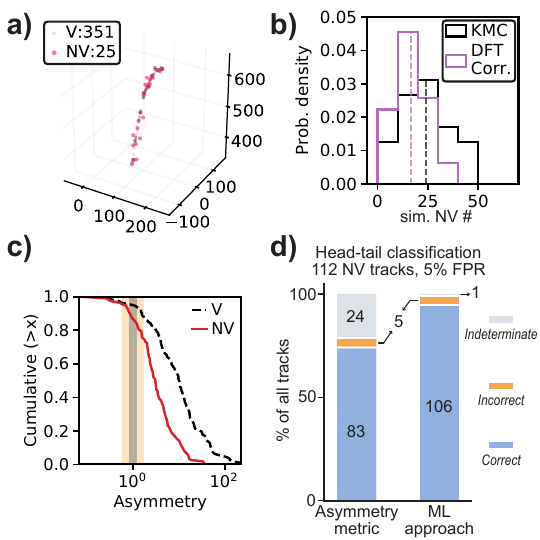}
    \caption{Modeling NV formation and head-tail directional information from simulated damage tracks. a) Example damage track before and after annealing. Prior to annealing, vacancies (V) generated by an 800\,keV carbon ion implantation are shown. During annealing, vacancies are consumed through recombination with interstitials, aggregation into vacancy clusters, or capture by substitutional nitrogen (not shown) to form NV centers. The resulting NV distribution after annealing is shown. b) Histogram of the number of NV centers per simulated implantation event. A mean NV number of $23.9\pm1.1$ (standard error of mean; standard deviation: 11.5) is obtained from 112 KMC-annealed tracks. A correction accounting for second-nearest-neighbor (2NN) defect interactions (DFT Corr.) reduces the predicted yield to $16.6\pm0.7$ NVs per event, in close agreement with experiment (Fig. \ref{fig:confocalscanres}d). c) Cumulative distribution of the asymmetry metric for vacancy (before annealing) and NV (after annealing) tracks. The asymmetry is defined as the ratio of defect counts in the tail and head regions of the track. For a threshold $x\geq1$, the shaded region between $1/x$ and $x$ indicates the indeterminate range. Tracks to the left of the shaded region are incorrectly classified and counted as false positives, while tracks to the right are correctly classified and counted toward the classification efficiency, following Ref. \cite{rajendran_method_2017}. For both vacancy and NV tracks, the thresholds are chosen to give a $\lesssim$5\% false positive rate. The resulting classification efficiency decreases from 93\% for vacancy tracks to 74\% for NV tracks, due to annealing-induced diffusion and reduced NV yield. d) Comparison of head-tail classification performance using NV tracks with the counting-based asymmetry metric and a machine learning (ML) approach. The ML model outputs a posterior probability over track direction, allowing tracks to be similarly classified as correct, incorrect, or indeterminate (see main text). At the same $\lesssim$5\% false positive rate, the ML method improves classification efficiency from 74\% (asymmetry metric) to 95\%, demonstrating enhanced recovery of directional information from NV spatial distributions.} 
    \label{fig:siimplkmc}
\end{figure}

To understand the measured NV yield, we developed a simulation pipeline that models both implantation and annealing processes. The implantation stage uses SIIMPL \cite{janson2003hydrogen}, which incorporates binary collision processes and the crystal lattice structure to produce spatial distributions of vacancies and interstitials constituting the damage track. These defect configurations are then used as input for kinetic Monte Carlo (KMC) simulations of annealing performed in SPPARKS \cite{mitchell2023spparks}, where diffusion and reactions among vacancies, interstitials, and substitutional nitrogen impurities are considered (see Methods). This framework predicts both the number and spatial distribution of NV centers following annealing (Fig. \ref{fig:siimplkmc}a). 
We note that the KMC simulation does not model the charge state of NV centers after formation. Because our experiment primarily measures NV$^-$ PL, direct comparison with simulation assumes that the NV$^0$ population does not significantly affect the inferred NV count, consistent with expectations for nitrogen-rich HPHT diamond \cite{deakFormationNVCenters2014,manson2018}.
The distribution of NV counts per simulated implantation event is shown in Fig. \ref{fig:siimplkmc}b. Across 112 simulated damage tracks, approximately 328 vacancies are produced per track on average. After annealing, a mean of $N_{\rm NV}$ = 23.9 NV centers is predicted (standard error of the mean: 1.1; standard deviation: 11.5), corresponding to a conversion yield of $\sim$7\%. As factors such as lattice displacement energy uncertainty and nitrogen concentration variation are too small to explain this $\sim$25\% difference from the experimental $N_{\rm NV}$ of 17.5 (see Supplementary Information), we investigated whether certain assumptions in our KMC simulation could account for the residual difference.

In the present KMC simulation, vacancy-related interactions, including vacancy-interstitial recombination, divacancy formation, and NV creation, are restricted to first-nearest-neighbor (1NN) separation. In this treatment, defects initially separated by second-nearest-neighbor (2NN) distances must first undergo isolated defect hopping to reach 1NN separation before reacting. To assess whether this restriction is physically justified, we performed nudged elastic band (NEB) calculations within density functional theory to perform a more precise determination of defect reaction barriers \cite{sheppard2012generalized} (see Supplementary Information). These calculations found that vacancy-related defect interactions can occur directly from 2NN configurations with little or strongly reduced energy barriers. Thus, vacancies with 2NN defect partners are expected to react effectively immediately, rather than first undergoing isolated defect hopping to reach 1NN contact.
This missing 2NN interaction range can affect the predicted NV yield in two ways. First, it reduces the initial population of free vacancies available for diffusion and later NV formation by removing vacancies that would react immediately with nearby interstitial or vacancy partners. Second, it may also modify the subsequent free-vacancy-to-NV conversion efficiency during annealing. Lastly, while vacancies initially adjacent to 2NN nitrogen will result in immediate NV formation, the number of such vacancies is negligible (see Supplementary Information). 

A full treatment of this effect would require implementing extended interaction ranges directly within the KMC framework. Doing so natively in SPPARKS is computationally expensive and non-trivial, as it requires replacing the lattice-local nearest-neighbor event catalog with distance- and configuration-dependent reaction rates; we therefore do not attempt a full 2NN-resolved KMC simulation here. Instead, as a first-order correction estimate, we assume that the dominant effect is the reduction of the initial free-vacancy population, and correct this population by excluding vacancies that have either 1NN or 2NN defect partners in the initial configuration. Rescaling this reduced population using the free-vacancy-to-NV conversion factor obtained from the original KMC simulation reduces the predicted mean NV yield to $16.6\pm0.7$ per simulated damage track, bringing the model into close agreement with experiment (Fig. \ref{fig:siimplkmc}b). We note that this estimate is approximate. Nonetheless, the analysis identifies short-range defect reactions beyond the 1NN assumption as the dominant missing physics. Additional factors that could influence NV yield are discussed in the Supplementary Information.

\subsection{Directional information from NV distributions}
Using the spatial distributions of defects before and after annealing from the simulation model, we investigated how directional information is modified by annealing. In contrast to vacancies, NV centers provide a viable optical probe of the damage track, as they are bright and observable down to the single-defect level. We evaluated head-tail directional information using a counting-based asymmetry metric following Ref. \cite{rajendran_method_2017}, where the asymmetry $A$, defined as the ratio of defect counts between the tail and head regions of the track, encodes the initial recoil direction because the tail region typically exhibits higher energy deposition (Bragg-peak-like behavior).
To benchmark classification performance, we choose an asymmetry threshold $x\geq1$. Tracks with $A>x$ are classified along the correct recoil direction, tracks with $0<A<1/x$ are classified in the opposite direction and counted as false positives, and tracks with $1/x\leq A\leq x$ are treated as indeterminate. We define the classification efficiency $\epsilon$ as the fraction of all tracks that are correctly classified, and the false positive rate (FPR) as the fraction of all tracks that are incorrectly classified. An effective detection scheme should therefore achieve high efficiency at low FPR. As shown in Fig. \ref{fig:siimplkmc}c, choosing a $\lesssim$5\% FPR for simulated 800\,keV tracks gives an efficiency of 93\% for vacancy tracks. After annealing, due to diffusion-induced broadening of the NV spatial distribution and reduced NV yield, the efficiency computed from NV tracks drops to 74\%.

We next tested whether a machine learning (ML) approach could recover directional information in NV tracks beyond the counting-based asymmetry metric. The model is based on simulation-based inference (SBI) \cite{tejero2020sbi}, in which a neural network learns an approximate posterior over recoil parameters given an observed defect distribution. This approach can exploit spatial features beyond simple count asymmetry; for example, the tail region of a track typically exhibits higher defect density, leading to reduced spacing between NV centers that is not captured by counting-based metrics.
To enable efficient training and improve generalization, each track was encoded as a low-dimensional vector of summary features describing the longitudinal defect distribution, including binned depth profiles, asymmetry measures, and the spread of defects on either side of the track centroid (see Methods and Supplementary Information). The neural network was trained on simulated vacancy tracks, which can be efficiently generated using SIIMPL, and then applied to NV distributions obtained after KMC annealing. 
The model outputs a probability ($P$) for the recoil direction. Analogous to the asymmetry-threshold analysis, we choose a probability threshold $y\geq0.5$, where tracks with $P>y$ are classified as correct, tracks with $P<1-y$ are counted as false positives, and tracks with $1-y \leq P \leq y$ are treated as indeterminate. At a comparable $\lesssim$5\% FPR, the ML approach yields a classification efficiency of 95\%, a $\sim$20 percentage-point increase over the counting-based asymmetry metric (Fig. \ref{fig:siimplkmc}d). Additional gains may be achieved with models trained directly on NV distributions, though this would require substantially larger KMC annealing datasets. Overall, these results demonstrate that data-driven approaches can recover directional information beyond simple counting-based metrics, and extending them to full vectorial direction reconstruction remains an important next step.

\subsection{Spin coherence of ion-implanted NV centers in diamond}

\begin{figure}[htbp]
    \centering
    \includegraphics[width=\textwidth]{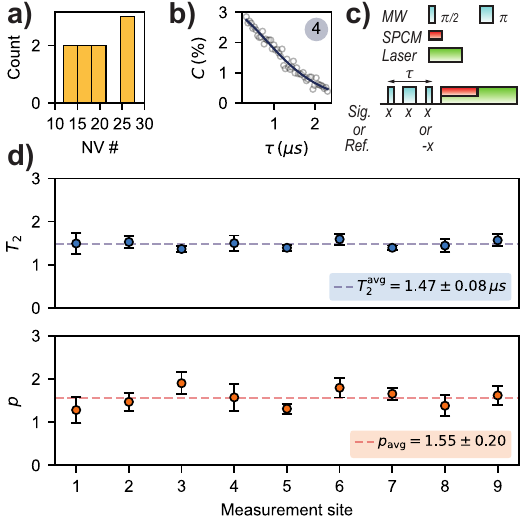}
    \caption{Spin coherence of ion-implanted NV centers. a) Distribution of NV numbers in the implantation sites selected for coherence measurements. b) Representative Hahn echo measurement (site 4). The signal is fitted to $C_0 \exp[-(\tau/T_2)^p]$ to extract the coherence time $T_2$ and decay exponent $p$. c) Hahn echo pulse sequence used for the measurements. The phase of the final microwave $\pi/2$ pulse is alternated between signal and reference shots to obtain the NV PL contrast $C$. d) Extracted coherence time $T_2$ (top) and decay exponent $p$ (bottom) for nine single ion implantation sites. Error bars represent the standard deviation from fitting. The measured average values ($T_2^{\rm{avg}}\approx1.47\,\mu\mathrm{s}$, $p_{\rm{avg}}\approx1.55$) are consistent with nitrogen spin-bath-limited ensemble decoherence. } 
    \label{fig:t2}
\end{figure}

Having established that NV distributions retain directional information after annealing, we next assessed the feasibility of resolving individual NV centers using super-resolution imaging. Magnetic gradient-based imaging has emerged as a leading super-resolution approach, enabling few-nanometer spatial resolution of multiple NV centers \cite{amawiThreedimensionalMagneticResonance2024}. The achievable resolution in such techniques depends on both the applied magnetic field gradient and the spin coherence properties of the NV centers. We therefore measured the ensemble NV spin coherence time across implantation tracks using a Hahn echo pulse sequence (Fig. \ref{fig:t2}a-c), extracting the coherence decay time $T_2$ and the decay exponent $p$, which encodes the local spin environment~\cite{Bauch2020}. As shown in Fig. \ref{fig:t2}d, we observe an average $T_2\approx1.47\,\mu$s, consistent with a nitrogen spin-bath-limited regime at the measured nitrogen impurity concentration ($[\mathrm{N}_{\rm{s}}^0]\approx253\,\mathrm{ppm}$) \cite{bauch2018ultralong}. Following Ref.~\cite{Bauch2020}, the extracted exponent $p\approx1.55$ indicates ensemble NV decoherence behavior.

These coherence properties enable estimation of the achievable spatial resolution for gradient-based imaging of NV distributions. Using a modest magnetic field gradient of 
$0.002\,\mathrm{G/nm}$ as demonstrated in Ref. \cite{amawiThreedimensionalMagneticResonance2024}, the measured coherence time corresponds to a spatial resolution of ${\sim}30\,\mathrm{nm}$. With state-of-the-art gradient strength $>0.01\,\mathrm{G/nm}$~\cite{zhang_selective_2017}, this resolution can be further improved to $<6\,\mathrm{nm}$, and for the 800\,keV implantation energy considered here, the average NV-NV separation is ${\sim}12\,\mathrm{nm}$.
These results establish the feasibility of NV-based nanoscale readout for reconstruction of track morphology in HPHT diamond.

\section{Discussion}

We have shown that for nuclear recoil energies of 800\,keV, individual recoil events in HPHT diamond can be detected through the formation of NV centers. We have also developed a simulation framework, calibrated to this experiment, that predicts directional information is retained in the post-annealing NV distribution. These implantation-induced NV centers also exhibit spin-echo coherence times consistent with the resolution requirements for nanoscale reconstruction of damage tracks using magnetic gradient-based quantum sensing. We emphasize that the present work demonstrates micrometer-scale localization of individual recoil events and quantifies the NV-formation process; nanoscale imaging of the NV distribution within a single damage track, required for full head-tail reconstruction from measured (rather than simulated) NV positions, remains an essential next step that will leverage the spin-coherence properties characterized here. Together, these results establish a viable pathway toward NV-based readout for directional detection in HPHT diamond at high recoil energies, with potential applications in the detection of energetic neutrons, neutrinos, and ions.

This framework can be extended to assess the applicability of NV-based directional detection in HPHT diamond at lower recoil energies. At 100\,keV, we performed annealing simulation for 84 tracks. The model predicts a mean of $13.8\pm0.7$ NV centers per track, formed from an average of 217 initial vacancies, corresponding to a yield of $\sim$6\%, similar to the 800\,keV case. However, the reduced number of NV defects lowers head-tail discrimination efficiency to 37\% using the counting-based asymmetry metrics at $\lesssim$5\% FPR. Applying the same ML analysis improves the efficiency to 96\% at the same FPR (see Supplementary Information). At recoil energies of $\sim$10\,keV, assuming a similar yield ratio, an average of $\sim$4 NV centers is expected to form from $\sim$50 vacancies, making directionality reconstruction based solely on NV positions challenging. Nevertheless, the NV PL signal remains above background ($\lesssim2$ NVs per confocal spot), enabling event localization at the micron scale.

These results suggest that a hybrid detection strategy will be required at low recoil energies. While NV centers provide a viable optical probe for event localization, a significant fraction ($\sim$40\%) of vacancies remain after annealing in the form of divacancies or higher-order defect clusters (with $\sim$50\% recombining with interstitials). These remaining vacancy-related defects generate local strain fields that may be detectable using high-resolution X-ray diffraction techniques~\cite{holtNanoscaleHardXRay2013}. Based on a previous study, it is estimated that for a typical 10\,keV recoil, strain fields on the order of $2\times10^{-4}$ can be generated within $\sim$30\,nm of the damage track, which is within experimentally accessible limits \cite{marshall_xray_2021}. Accounting for vacancy recombination during annealing reduces this estimate by approximately a factor of two, but the resulting strain fields remain measurable. More detailed modeling of strain fields from annealed defect configurations will be required for quantitative predictions. Overall, these considerations indicate that combining NV-based readout with nanoscale strain-based imaging provides a promising pathway for investigating directional detection of low-energy recoil events.

The framework developed here may also have broader implications beyond directional detection. In the context of paleodetectors for DM searches, track preservation over geological timescales is often assumed; however, the strong exponential temperature dependence of defect diffusion \cite{fleischer1965solid} suggests that transient periods of elevated temperature may modify or erase track signatures. Incorporating thermal histories into defect evolution models may therefore be important for interpreting candidate materials such as olivine. In addition, this approach can be extended to model implantation and annealing processes in engineered quantum materials. For example, incorporating surface boundaries or non-uniform dopant distributions, such as delta-doped nitrogen layers \cite{kim2025scalable}, would enable exploration of parameter spaces including implantation energy, dose, and angle, and their impact on NV formation and the resulting lattice environment.

\section{Methods}
\subsection{Sample preparation and ion implantation}

A high-pressure, high-temperature (HPHT) type Ib single-crystal diamond sample (Element Six) with a specified nitrogen concentration of $[\mathrm{N}_{\rm{s}}^0]\sim200\,\mathrm{ppm}$ and $\{100\}$ surface orientation was used. The sample surface was polished to a roughness of $R_a<1\,\mathrm{nm}$, and fiducial markers were fabricated to guide implantation and subsequent confocal optical characterization. 

Carbon ion implantation at 800\,keV was performed using the 6\,MV Tandem accelerator at the MicroOne endstation at Sandia National Laboratories \cite{Chu:2006,MARTIN1981}. The ion beam had a full width at half maximum (FWHM) spot size of 2.62\,$\mu$m (X) $\times$ 2.90\,$\mu$m (Y), and implantation was carried out on a grid of sites with spacings of 6\,$\mu$m or 10\,$\mu$m. The beam current and dwell time were adjusted to achieve an average dose of about one ion per site, maximizing the probability of single-ion implantation events. To minimize ion channeling effects, the sample surface was tilted by ${\sim}11\degree$  relative to the incident ion beam direction \cite{nordlund_channeling_2016}. 

Annealing was performed in a vacuum tube furnace (MTI OTF-1200X) at 800\degree C and $10^{-6}$ torr for 2 hours, following standard recipes~\cite{pezzagna_creation_2010,SangtawesinDeLeon2019surfaceprep}. NV center formation was confirmed by laser-induced PL (see below). To verify that annealing had proceeded to completion, a second identical cycle was performed; the absence of any detectable change in PL confirmed that the initial conditions were sufficient.

\subsection{Confocal microscopy and NV measurement}

A home-built confocal microscope was used to characterize the diamond samples. NV centers were optically excited using a 532\,nm laser (Coherent Compass 315M) with a typical power of ${\sim}1\,\mathrm{mW}$ at the sample. The excitation beam was circularly polarized to uniformly excite NV centers of different crystallographic orientations. NV PL in the 647-800\,nm range was collected onto a single-photon counting module (Excelitas SPCM-AQRH-14). 

The collimated excitation beam ($d\approx3.5\,\mathrm{mm}$) was focused through a 100×, 0.8 NA objective (Nikon CFI60 TU Plan Epi ELWD), and collected PL was spatially filtered using a $100\,\mu \mathrm{m}$ pinhole to reject out-of-focus light. The microscope operated in a sample-scanning configuration using a nanopositioning stage (Mad City Labs Nano-3D200). The point spread function (PSF) was characterized by three-dimensional photoluminescence scans of a single NV center, yielding Gaussian widths of $\sigma_x=189\pm12\,\mathrm{nm}$, $\sigma_y=180\pm11\,\mathrm{nm}$, and $\sigma_z=546\pm26\,\mathrm{nm}$.

For spin measurements, microwave (MW) control was provided by a continuous-wave source (SRS SG384), gated using a fast electronic switch driven by a pulse generator (Swabian Pulse Streamer 8/2). The same pulse generator synchronized MW pulses with optical excitation via an acousto-optic modulator (Brimrose TEM-85-10) to implement pulsed sequences such as Hahn echo. NV photoluminescence counts were recorded using a data acquisition system (NI USB-6363) for spin-state analysis.

\subsection{NV formation simulation}
Ion implantation was modeled using the SIIMPL (Simulation of ion IMPLantation) binary collision approximation code \cite{janson2003hydrogen,siimpl}, which was chosen over the more commonly used SRIM/TRIM code~\cite{ZieglerSRIM2010} to allow modelling the diamond crystalline structure and ion channeling effects, unlike SRIM which assumes amorphous targets. The simulation was run at 800\,keV carbon ion energy and 11\degree~polar angle (reflecting the Sandia implantation parameters), with lattice displacement threshold energy and thermal vibration amplitude set to the standard values of 43\,eV~\cite{koike1992displacement} and 0.044\,\AA~\cite{beyerElectronDensityThermal2023} respectively. The simulation produced lists of vacancy and interstitial coordinates for each implanted ion. 

Annealing was modeled using a kinetic Monte Carlo (KMC) simulation implemented in SPPARKS~\cite{mitchell2023spparks}, which tracks defect diffusion and reactions on the diamond lattice. Transition rates were described using an Arrhenius form, $\gamma=\nu_0 \exp(-\frac{E_a}{k_BT})$, at an annealing temperature of \SI{800}{\celsius}. Initial defect configurations of vacancies and interstitials were obtained from SIIMPL implantation simulations. To improve computational efficiency, damage tracks were partitioned into spatially separated clusters, which were treated as independent (spacing $\gtrsim154\,\mathrm{\AA}$, interaction probability $\lesssim1$\%; see Supplementary Information). For each cluster, a simulation domain was constructed by embedding the defect configuration within a diamond lattice padded by 100 unit cells in all directions, ensuring negligible vacancy loss through boundary diffusion. Nitrogen atoms were randomly distributed at a concentration of 253\,ppm, consistent with experimental measurements (see Supplementary Information). Results from individual clusters were combined to obtain statistics for full damage tracks. Typical simulations contained on the order of $10^8-10^9$ atoms per cluster.

Defect diffusion and reactions were modeled as follows. Vacancies were allowed to hop with an activation energy of $E_a=2.59$\,eV \cite{deakFormationNVCenters2014,Alsid2019nvcreation}, while carbon interstitials were assigned $E_a=1.5$\,eV \cite{breuer1995ab}. All other defects, including substitutional nitrogen atoms, were assumed to be immobile. Reactions included vacancy recombination with interstitials (restoring the lattice), vacancy aggregation into divacancies and higher-order clusters ($\mathrm{V+V\xrightarrow{}VV}$, $\mathrm{V+VV\xrightarrow{}VVV}$), and NV center formation via vacancy capture by nitrogen ($\mathrm{V+N\xrightarrow{}NV}$). These reactions were assumed to occur upon first-nearest-neighbor contact, corresponding to effectively barrierless interactions.

\subsection{Head-tail directionality and machine learning (ML)}

Head-tail directionality was first evaluated using an asymmetry metric $A$, defined as the ratio of defect counts between the tail (top one-third) and head (bottom one-third) regions of a track. Exchanging the head and tail labels maps $A\rightarrow1/A$. For the simulated benchmark tracks considered here, we define $+z$ to be along the recoil direction, such that the tail is more positive in $z$ than the head. Therefore, given a threshold $x\geq1$, tracks with $A>x$ were classified as correctly identified, tracks with $A<1/x$ were classified as incorrectly identified, and tracks with $1/x\leq A\leq x$ were treated as indeterminate.

For the ML analysis, the model outputs a posterior probability $P$ for the recoil direction. In our convention, $P\rightarrow1$ corresponds to motion along $+z$, while $P\rightarrow0$ corresponds to motion along $-z$. For test tracks with ground-truth recoil direction along $+z$, we applied a threshold $y\geq0.5$: tracks with $P>y$ were classified as correct, tracks with $P<1-y$ were classified as incorrect, and tracks with $1-y\leq P\leq y$ were treated as indeterminate. For both the asymmetry and ML analyses, the threshold was chosen to evaluate the classification efficiency at a fixed false positive rate. More details of the ML analysis can be found in the Supplementary Information.

\section{Acknowledgements}

We thank Reza Ebadi and Mason Marshall for their contributions towards early stages of this work; Mason Camp, Johannes Cremer, and Connor Hart for useful discussions; and Gajadhar Joshi for sample preparation assistance at Sandia. This work was supported by, or in part by, the Argonne National Laboratory under Award No. 2F60042; the DOE fusion program under Award No. DESC0021654; the U.S. Army Research Laboratory under Contract Nos. W911NF1920181 and W911NF2420143; and the University of Maryland Quantum Technology Center. ToF-SIMS analysis was carried out with support provided by the National Science Foundation CBET-1626418. This work is conducted in part using resources of the Shared Equipment Authority at Rice University. VI and JC acknowledge support from the National Science Foundation Growing Convergence Research award 2428507, and the DARPA QuSen Program. First-principles calculations were carried out with computational resources and technical support provided by Advanced Research Computing at Virginia Tech (arc.vt.edu). For the annealing simulations, the authors acknowledge the University of Maryland supercomputing resources, with additional computations performed at the Advanced Research Computing at Hopkins (ARCH) core facility, supported by NSF grant OAC-1920103; we thank Surjeet Rajendran (Johns Hopkins University) for generously sharing allocation time on the Rockfish cluster.

\section*{Author contributions}
D.G.A. and J.T. contributed equally to this work. D.G.A. and J.T. conceived and designed the study and performed the confocal and NV spin measurements and data analysis. M.T. and E.S.B. performed the ion implantation at Sandia National Laboratories. T.T. performed the ToF-SIMS measurements. V.I. and J.C. carried out the first-principles (DFT/NEB) calculations. J.T. developed the SIIMPL and SPPARKS simulation pipeline. C.B. developed the machine-learning analysis. M.S., G.L., and A.T.G. contributed to experimental measurements, simulations, and data analysis. D.G.A. and R.L.W. supervised the project. D.G.A., J.T., and R.L.W. wrote and edited the manuscript with input from all authors. All authors reviewed the manuscript. 

\section*{Competing interests}
The authors declare no competing interests.

\section*{Data availability}
The data that support the findings of this study are available from the corresponding authors upon reasonable request.

\section*{Code availability}
The ion-implantation simulations used the open-source SIIMPL code~\cite{siimpl}; the annealing simulations used the open-source SPPARKS package~\cite{mitchell2023spparks}; and the machine-learning analysis used the open-source \texttt{sbi} toolkit~\cite{tejero2020sbi}. Custom code for damage-track cluster decomposition, NV yield extraction, and head-tail classification is available from the corresponding authors upon reasonable request.

   \clearpage
   \setcounter{section}{0}
   \setcounter{figure}{0}
   \setcounter{table}{0}
   \renewcommand{\thesection}{\Alph{section}}
   \renewcommand{\thefigure}{S\arabic{figure}}
   \renewcommand{\thetable}{S\arabic{table}}
   \section*{Supplementary Information}

\section{NV photoluminescence (PL) analysis details}
\label{si:nv}

\subsection{Data analysis procedure for confocal scans}

\begin{figure}[htbp]
    \centering
    \includegraphics[width=\linewidth]{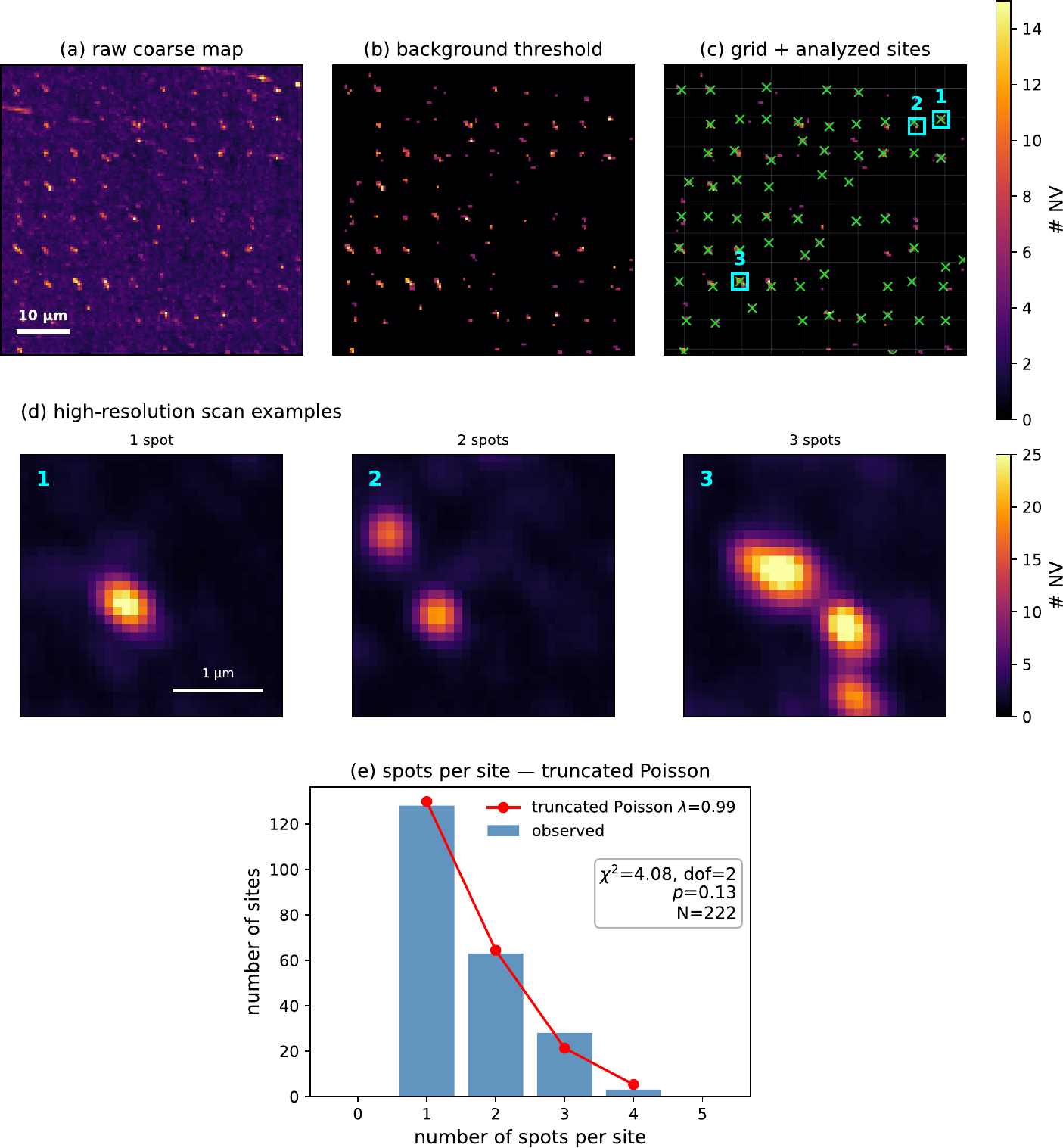}
    \caption{\textbf{NV-count extraction pipeline and multi-ion statistics}, illustrated for a representative 800\,keV implantation square (6\,$\mu$m pitch). \textbf{(a)} Raw coarse confocal map (500\,nm pixels), in units of calibrated single-NV PL. \textbf{(b)} After region-of-interest and scratch masking and intensity thresholding, isolating the implantation-site PL. \textbf{(c)} Elliptical grid mask ($\pm3\sigma$ of the beam profile) overlaid on the thresholded map; green crosses mark the analyzed sites and cyan boxes the three example sites shown in (d). \textbf{(d)} High-resolution nanovolume scans of the three boxed sites, resolving one, two, and three PL spots. \textbf{(e)} Distribution of the number of resolved spots per site over all $N = 222$ analyzed 800\,keV sites (bars), with a maximum-likelihood truncated-Poisson fit (red; $\lambda = 0.99$, $\chi^2 = 4.1$, $2$ d.o.f., $p = 0.13$).}
    \label{fig:si_pipeline}
\end{figure}

The number of NV centers formed at each implantation site is extracted from the confocal PL data in several stages (Fig.~\ref{fig:si_pipeline}). First, 100\,$\mu$m $\times$ 100\,$\mu$m implantation squares on the sample containing a regular grid of sites at 6 or 10\,$\mu$m pitch are imaged in a coarse, wide-field two-dimensional scan at 500\,nm pixel pitch, focused at $\sim$600\,nm below the surface (Fig.~\ref{fig:si_pipeline}a). The following procedure is then used to locate candidate implantation sites for finer scans. PL count rates are converted to a number of NV centers using a single-NV reference rate $I_1 = 25$\,kcts/s, calibrated on isolated NV centers in a low-background region of the same diamond sample under identical excitation and collection conditions. Sub-regions within implantation squares containing high NV backgrounds from residual subsurface polishing damage are manually excluded, and an intensity window of $5$--$100\,I_1$ ($125$--$2500$\,kcts/s) is applied to suppress the residual background ($\lesssim2$ NV per confocal spot; Fig.~\ref{fig:si_pipeline}b).\footnote{See section \ref{sec:diamondsurfaceprep} for results of additional diamond surface preparation experiments to suppress such backgrounds for future implantations.} 

Next, a rectangular grid of points spaced by the nominal implantation pitch (6 or 10\,$\mu$m) is overlaid on the image, and an elliptical acceptance region with semi-axes equal to $\pm3\sigma$ of the measured beam profile is placed at each grid node. Within each ellipse, the brightest pixel $\mathbf{r}_i^{\mathrm{peak}}$ is taken to mark the implantation site. The grid origin is then shifted by an offset $(\delta_x,\delta_y)$ chosen to minimize the total distance between the grid nodes and their enclosed brightest pixels,
\begin{equation}
D(\delta_x,\delta_y) = \sum_i \bigl\lVert \mathbf{r}_i^{\mathrm{grid}}(\delta_x,\delta_y) - \mathbf{r}_i^{\mathrm{peak}} \bigr\rVert ,
\label{eq:gridoffset}
\end{equation}
where $\mathbf{r}_i^{\mathrm{grid}}$ is the position of the $i$-th grid node, which shifts with the applied offset $(\delta_x,\delta_y)$. Grid nodes whose brightest enclosed pixel exceeds $25\,I_1$ are excluded from the sum as anomalously bright artifacts (Fig.~\ref{fig:si_pipeline}c). This procedure is similar to that of Ref.~\cite{yamamotoIsotopicIdentificationEngineered2014}.

From the procedure above, each located site is matched to a high-resolution three-dimensional confocal scan ($\sim$200\,nm pixel pitch) acquired at that position, from which the NV number is obtained as $N_{\rm NV} = (I_{\rm max} - I_{\rm bg})/I_1$, where $I_{\rm max}$ is the brightest pixel of the in-focus slice and $I_{\rm bg}$ the median of the scan's border pixels (Fig.~\ref{fig:si_pipeline}d). Sites with one to six resolved PL spots are retained, and per-site outliers are removed if the NV count deviates from the population median by more than three times the rescaled median absolute deviation ($1.4826\times\mathrm{MAD}$, which estimates the standard deviation for normally distributed data). Across the 800\,keV region the procedure yields $N = 222$ analyzed sites with a mean NV yield $\langle N_{\rm NV}\rangle = 17.5 \pm 0.38$ (standard error of the mean, $5.7/\sqrt{N}$) and a site-to-site standard deviation of $5.7$ (main text Fig.~\ref{fig:confocalscanres}d).

The reported mean NV count is robust to the analysis choices. Tests were performed varying the grid offset, elliptical acceptance area, outlier rejection brightness threshold, and the per-site intensity estimator (brightest pixel versus the mean of the ten brightest pixels), as well as restricting the data set to sites with a single instead of multiple resolved spots. None of these data analysis variations resulted in significant changes to $\langle N_{\rm NV}\rangle$. 

\subsection{Poisson analysis of multi-ion impacts}
At sites that received more than one ion, the high-resolution scans resolve multiple, spatially separated PL spots (Fig.~\ref{fig:si_pipeline}d), each attributed to a single ion impact. The distribution of the number of resolved spots per site, pooled over all $222$ analyzed sites, is shown in Fig.~\ref{fig:si_pipeline}e. Because sites with zero spots are not observed, we model the counts with a Poisson distribution truncated to $n \ge 1$, \begin{equation} P(n) = \frac{\lambda^{n} e^{-\lambda}}{n!\,\bigl(1 - e^{-\lambda}\bigr)}, \qquad n = 1, 2, 3, \dots, \end{equation} whose single parameter $\lambda$ is the mean number of ions delivered per grid node. A maximum-likelihood fit gives $\lambda = 0.99$. A Pearson $\chi^2$ goodness-of-fit test, with the counts grouped into bins $n = 1$, $2$, $3$, and $\ge 4$ so that every expected count exceeds five, yields $\chi^2 = 4.1$ for $2$ degrees of freedom ($p = 0.13$); the data are thus statistically consistent with a truncated Poisson distribution. This confirms that each resolved PL spot corresponds to a single, independent ion impact, since the arrival of ions at a given grid node is expected to follow Poisson statistics. 

\section{SIMS measurement of nitrogen concentration}
The kinetic Monte Carlo (KMC) annealing simulations described in the main text take the substitutional nitrogen concentration $[\mathrm{N}_{\rm{s}}^0]$ as an input. We determined this value for the implanted sample (``Sandia 1") using two independent methods: Time-of-Flight Secondary Ion Mass Spectrometry (TOF-SIMS) and an estimate based on the NV background PL of the annealed sample. Both methods are anchored to a commercial SIMS measurement of a second HPHT diamond sample (``Sandia 2") from the same manufacturer (Element Six). We use the average of the two estimates as the simulation input, treating the spread between them as a conservative uncertainty range; as shown in Section~\ref{sec:add-factors-nv-yield} (Fig.~\ref{fig:sm_factorsNVyield_2}), the predicted NV yield is only weakly dependent on $[\mathrm{N}_{\rm{s}}^0]$ over this range, so the residual calibration uncertainty does not affect the conclusions of the main text.

\subsection{TOF-SIMS depth profiling}\label{sec:TOF-SIMS}
Negative high mass-resolution depth profiles were acquired with a TOF-SIMS NCS instrument, which combines a TOF.SIMS5 spectrometer (ION-TOF GmbH, Münster, Germany) and an in-situ scanning probe microscope (NanoScan, Switzerland) at the Shared Equipment Authority from Rice University. The analysis field of view was $80\times80\,\mu\rm{m}^2$ ($\rm{Bi}_3^+$ at 30\,keV, 0.3\,pA) with a raster of 256 by 256 along the depth profile. Charge compensation was provided by an electron flood gun, with surface potential adjusted to $20$\,V. The cycle time was $70\,\mu$s, corresponding to a mass range of $m/z = 0$--$911$\,amu. Sputtering was performed with Cs$^+$ at $2$\,keV, $105$\,nA, rastered over $100\times 100\,\mu\mathrm{m}^2$. The two beams were operated in non-interlaced mode, alternating one analysis cycle with ten sputtering frames followed by a $5$\,s charge-compensation pause. Depth calibration was performed using the interface tool in SurfaceLab v7.3 (ION-TOF GmbH), with sputter rates referenced to crater depths measured ex-situ with a DekTak stylus profilometer. For each region of interest, the analyzed area was further cropped to remove crater edges and ensure homogeneity of the extracted signal.

\begin{figure}[htbp]
    \centering
    \includegraphics[width=\textwidth]{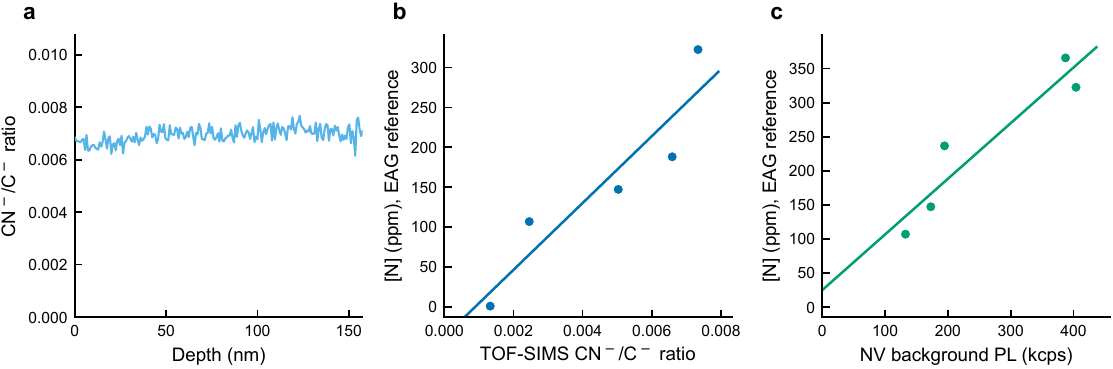}
    \caption{\textbf{Nitrogen-concentration measurement.}
        \textbf{(a)} Representative TOF-SIMS depth profile of the CN$^-$/C$^-$ ion ratio, showing a uniform signal over the profiled depth.
        \textbf{(b)} EAG--Rice TOF-SIMS calibration: CN$^-$/C$^-$ ratio measured at Rice University plotted against the EAG reference nitrogen concentration for five regions of Sandia~2 (solid line: linear fit).
        \textbf{(c)} NV background PL calibration: confocal PL intensity plotted against the EAG reference nitrogen concentration for the same five regions (solid line: linear fit).}
    \label{fig:N-calibration}
\end{figure}

Quantification of the CN$^-$ ion signal into an absolute nitrogen concentration requires an external calibration. To this end, we sent a second HPHT diamond with similar properties (Sandia 2) to EAG for commercial SIMS analysis at multiple regions of interest, and then measured the same regions on Sandia 2 with the Rice TOF-SIMS. The resulting linear EAG-Rice calibration curve (Fig. \ref{fig:N-calibration}b) was applied to all subsequent Rice TOF-SIMS depth profiles acquired on Sandia 1. A representative depth profile is shown in Fig. \ref{fig:N-calibration}a. This SIMS procedure was performed across six different regions on Sandia 1 adjacent to implanted regions, yielding an average value of $[\mathrm{N}_{\rm{s}}^0]_{\mathrm{SIMS}} \approx 318$\,ppm.

\subsection{NV background PL as an independent estimate}\label{sec:fluor-estimate}

As an independent cross-check, we exploit the correlation between substitutional nitrogen concentration and grown-in NV-related PL in HPHT type Ib diamond \cite{vanEnckevort1988}. NV background PL intensities were measured in implanted regions of Sandia 1 (i.e., the regions surrounding the NV clusters at the ion impact sites) and at multiple regions of the EAG-measured Sandia 2 sample, the latter providing calibration. Plotting confocal PL intensity against the corresponding EAG-reported $[\mathrm{N}_{\rm{s}}^0]$ values yields a linear correlation (Fig. \ref{fig:N-calibration}c) which confirms the validity of this proxy for our HPHT material. Applying this calibration to the averaged background PL in implanted regions of Sandia 1 gives $[\mathrm{N}_{\rm{s}}^0]_{\mathrm{PL}} \approx 188$\,ppm.

\subsection{Combined estimate and propagation into the simulations}\label{sec:N-combined}

The two methods yield estimates that differ by approximately $\pm 25\%$ from their mean: $\approx\! 318$\,ppm from TOF-SIMS and $\approx\! 188$\,ppm from NV background PL. We adopt the average, $[\mathrm{N}_{\rm{s}}^0] = 253$~ppm, as the nominal input to the KMC annealing simulations, and treat the spread between the two methods as a conservative uncertainty range of $\sim$180--320 ppm. As shown in Section \ref{sec:add-factors-nv-yield} (Fig. \ref{fig:sm_factorsNVyield_2}), the predicted NV yield varies only weakly with $[\mathrm{N}_{\rm{s}}^0]$ across this range and indeed across nearly two orders of magnitude in concentration. The dominant determinants of NV yield are instead the vacancy-interstitial and vacancy-vacancy reactions discussed in Section \ref{sec:dft}, and the calibration uncertainty on $[\mathrm{N}_{\rm{s}}^0]$ does not propagate appreciably into the conclusions of the main text.

\section{Annealing simulation details}

\subsection{Cluster decomposition and KMC framework}
The defect distribution -- consisting of both vacancies and carbon interstitials -- produced by ion implantation forms an extended but spatially discontinuous damage track. The defect–defect interaction probability decreases with initial separation $r$ (assuming defects are mobile), and in a three-dimensional continuous limit, this interaction (or “hitting”) probability is described as $P(r)=a_0/r$, where $a_0$ is the interaction radius. Based on this scaling, we partition each damage track into smaller clusters and simulate each cluster independently, effectively assuming negligible interaction between clusters. 

\begin{figure}[htbp]
    \centering
    \includegraphics[width=\textwidth]{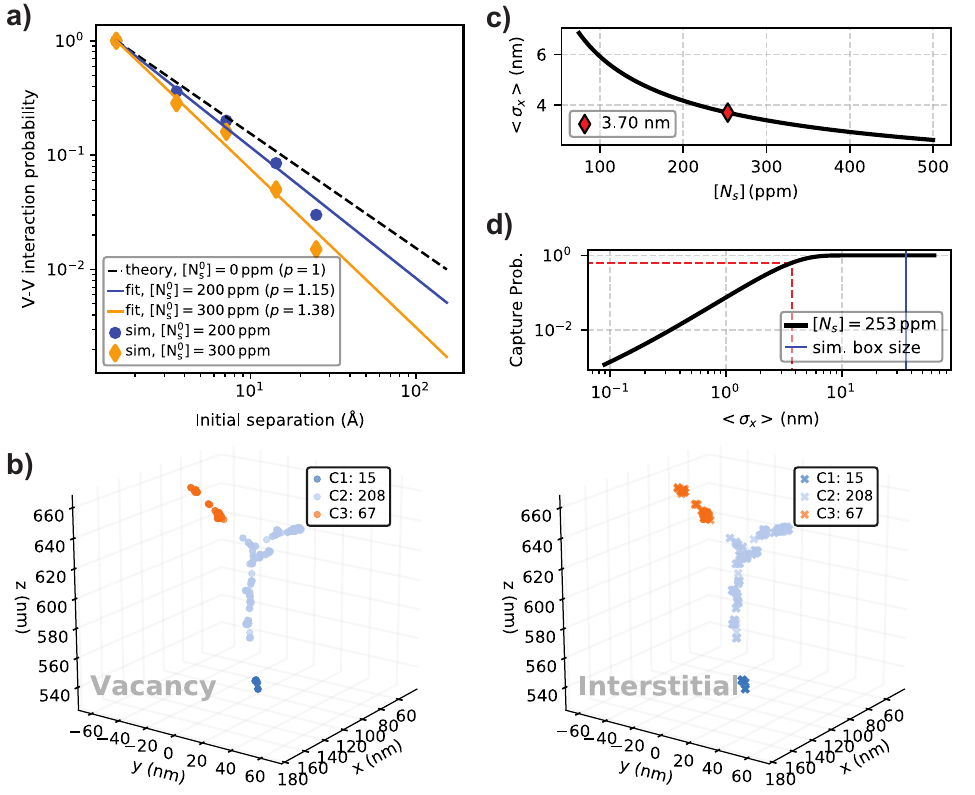}
    \caption{Cluster decomposition and lattice domain size selection for annealing simulations. (a) Probability of vacancy–vacancy interaction as a function of initial separation $r$, obtained from two-vacancy simulations at different nitrogen concentrations. The dashed theoretical line is computed using the 3D hitting probability, assuming an effective interaction radius of $1.54\,\text{\AA}$. Solid lines show power-law fits of the form $1.54\,\text{\AA}/r^p$. For the experimental nitrogen concentration  $[\rm{N}_{\rm{s}}^0]=253\,\rm{ppm}$, the interaction probability at $154\,\text{\AA}$ is $\sim$0.3\%, supporting negligible cross-cluster interactions. (b) Example decomposition of an 800\,keV implantation damage track into spatially separated clusters. The minimum pairwise defect spacing between clusters exceeds $154\,\text{\AA}$. Vacancy (left) and carbon interstitial (right) clusters are shown. The legend indicates the number of defects in each cluster. (c) Root-mean-square (RMS) displacement of a diffusing vacancy as a function of nitrogen concentration, evaluated at the point where the number of unique sites visited equals the number of available capture sites ($N_{uni}=N_{cap}$), corresponding to a capture probability of $\sim$64\%. (d) Vacancy capture probability as a function of RMS displacement for a nitrogen concentration of $[\rm{N}_{\rm{s}}^0]=253\,\rm{ppm}$. For each cluster, the simulation domain is defined by padding 100 unit cells in each direction from all vacancy defects. This ensures that the probability of vacancy escape is $\lesssim0.01$\%.} 
    \label{fig:sm_annealing_1}
\end{figure}

To ensure the validity of this approximation, defects belonging to different clusters are required to be separated by more than $154\,\text{\AA}$. Under the assumption that defect reactions occur at nearest-neighbor (1NN) separation (i.e. $a_0=1.54\,\text{\AA}$), this corresponds to an interaction probability of $\leq1$\% between any defect pairs from different clusters in the absence of other species. The presence of immobile substitutional nitrogen impurities can further suppress defect–defect interactions between clusters by providing an additional reaction channel. We quantify this effect for vacancy–vacancy interactions by directly simulating two-vacancy systems with varying initial separations $r$ and nitrogen concentrations $[\mathrm{N}_{\rm{s}}^0]$, as shown in Fig. \ref{fig:sm_annealing_1}(a). The interaction probability as a function of $r$ is fitted to a power-law dependence $1.54\,\text{\AA}/r^p$. For our experimentally measured nitrogen concentration of $[\mathrm{N}_{\rm{s}}^0]=253\,\mathrm{ppm}$, the expected vacancy-vacancy interaction probability at a separation of $154\,\text{\AA}$ is reduced to $\sim$0.3\%. In practice, vacancy–vacancy interactions and vacancy–interstitial recombination within each cluster further reduce the likelihood of cross-cluster interactions, supporting the validity of independent cluster simulations. An example of an 800\,keV damage track and its decomposition into clusters is shown in Fig. \ref{fig:sm_annealing_1}(b). The corresponding carbon interstitial distributions are also shown; due to their close spatial proximity to vacancies, interstitial clusters closely follow the vacancy distribution, with approximately equal numbers of vacancies and interstitials within each cluster.

After partitioning the damage track into clusters, a diamond lattice is constructed around each cluster. The simulation domain must be sufficiently large such that vacancies are unlikely to diffuse out of the box and instead undergo reactions within the domain. To ensure this condition, the box size is determined by requiring that trapping by substitutional nitrogen impurities alone is sufficient to capture diffusing vacancies.

For a vacancy undergoing a random walk on the diamond lattice, the number of unique lattice sites visited after $N$ hops is given by \cite{fastenau1982diffusion,Alsid2019nvcreation}
\begin{equation}
N_{\mathrm{uni}} \approx 0.56N + 0.63\sqrt{N}.
\end{equation}
For a given nitrogen concentration $[\rm{N}_{\rm{s}}^0]$ (in ppm), the expected number of distinct lattice sites that must be visited before encountering a nitrogen capture site is
\begin{equation}
N_{\mathrm{cap}} = \frac{10^6}{4[\mathrm{N}_s^0]},
\end{equation}
where the factor of 4 reflects the four nearest-neighbor sites adjacent to each substitutional nitrogen, each of which is a possible capture site for NV formation. The cumulative probability $\chi$ that a vacancy is captured by nitrogen after $N$ hops is 
\begin{equation}
\chi = 1 - \exp\left(-\frac{N_{\mathrm{uni}}}{N_{\mathrm{cap}}}\right),
\end{equation}
and the root-mean-square (RMS) displacement along any Cartesian direction is given by
\begin{equation}
\langle \sigma_{x/y/z} \rangle = \sqrt{\frac{N}{3}} \times 1.54\,\text{\AA}.
\end{equation}

Figure~\ref{fig:sm_annealing_1}(c) shows the RMS displacement as a function of nitrogen concentration at the point where $N_{uni}=N_{cap}$, corresponding to a capture probability of $\sim$64\%. To ensure that vacancies remain within the simulation domain, the lattice is padded by at least 100 unit cells in each direction from any vacancy defects in the simulated cluster. At this scale, the probability of a vacancy reaching the boundary without being captured is $\lesssim0.01$\%, as shown in Fig.~\ref{fig:sm_annealing_1}(d).

\begin{figure}[htbp]
    \centering
    \includegraphics[width=\textwidth]{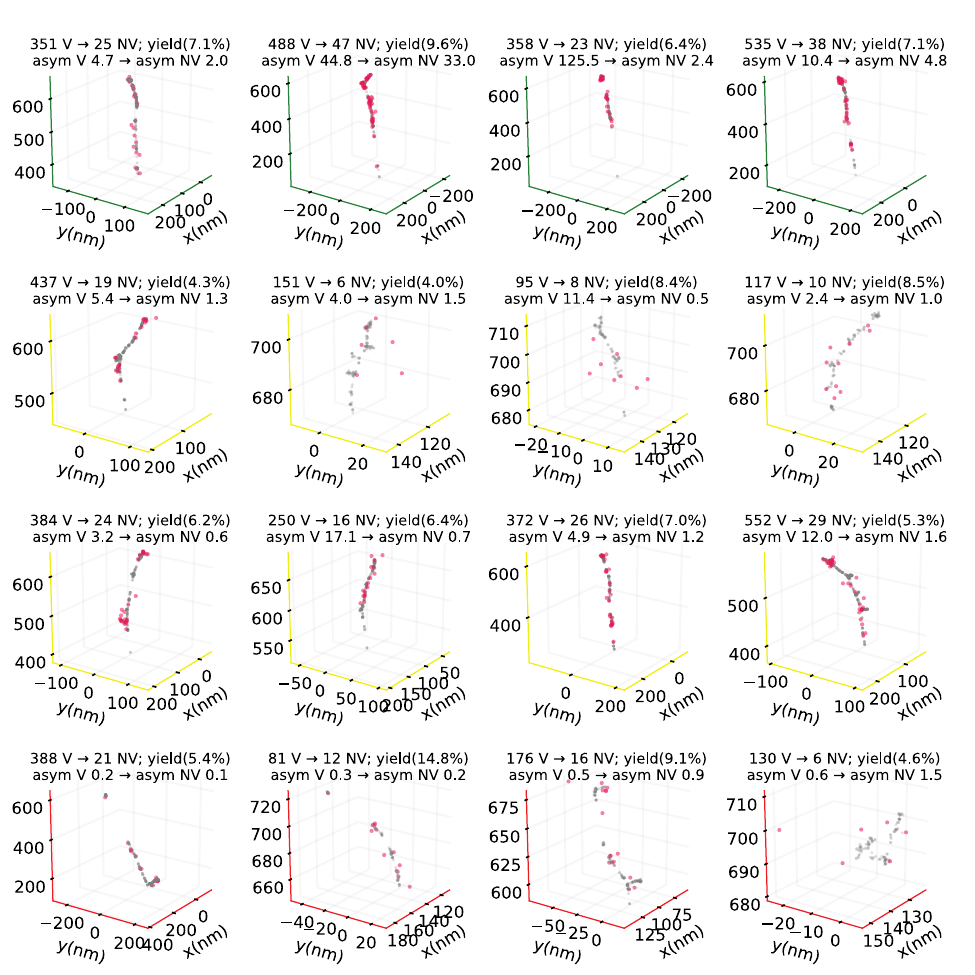}
    \caption{Representative damage tracks before and after annealing simulations. Gray dots indicate the initial vacancy distribution, and red dots indicate the resulting NV distribution after annealing. The vacancy-to-NV conversion yield and head-tail asymmetry values, computed using a defect counting method, are shown for both vacancy and NV tracks.} 
    \label{fig:sm_annealing_2}
\end{figure}

During the annealing simulation, the kinetic Monte Carlo (KMC) algorithm \cite{mitchell2023spparks} samples allowed processes -- including defect diffusion and reactions -- according to Arrhenius rates of the form $\gamma=\nu_0 \exp(-\frac{E_a}{k_BT})$, which determine the relative frequency of each event. Here, $\nu_0$ is taken as the Debye frequency of diamond \cite{Alsid2019nvcreation}, and the annealing temperature is set to \SI{800}{\celsius}. After each event is selected and applied, the global lattice configuration is updated and the simulation time advances, resulting in non-uniform time steps that depend on the selected processes. The simulation is terminated when no further events are available, ensuring that all diffusion and reaction processes are exhausted. Additional damage tracks before and after annealing are shown in Fig. \ref{fig:sm_annealing_2}, with corresponding head-tail asymmetry values computed using the defect counting method. Diffusion-induced broadening during annealing reduces the asymmetry of the NV distribution relative to the initial vacancy track.

\subsection{Additional factors influencing NV yield}
\label{sec:add-factors-nv-yield}
Here we consider two sources of uncertainty that may affect the predicted NV yield: the total vacancy production in the implantation simulation and the measured nitrogen concentration. As discussed below, neither is sufficient to account for the discrepancy of NV yield between model predictions and experimental measurements. This further supports the need for DFT-informed corrections to defect interactions as an important mechanism governing NV formation.

\begin{figure}[htbp]
    \centering
    \includegraphics[width=3.6in]{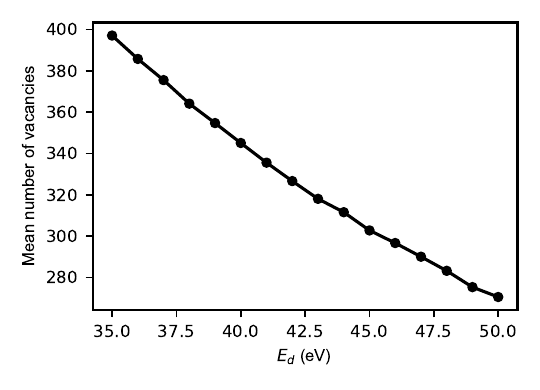}
    \caption{Sensitivity of vacancy production to displacement energy $E_d$. Mean number of vacancies generated per 800\,keV carbon recoil as a function of the threshold displacement energy used in SIIMPL simulations. Each data point represents the average over 5000 simulated damage tracks.} 
    \label{fig:sm_factorsNVyield_1}
\end{figure}

\textbf{Displacement energy.} The threshold displacement energy $E_d$ used in the SIIMPL simulation controls the number of vacancies generated in the damage track. In diamond, $E_d$ is known to be directionally dependent \cite{koike1992displacement}, with reported values ranging from $\sim$37.5\,eV for [100] to $\sim$47.6\,eV for [110]. Similar to SRIM, SIIMPL employs a single scalar value of $E_d$, effectively averaging over crystallographic directions during the recoil cascade. In our analysis, we use $E_d=43$\,eV. To evaluate the sensitivity of vacancy production to this parameter, we vary $E_d$ over a broad range. As shown in Fig.~\ref{fig:sm_factorsNVyield_1}, the mean vacancy number decreases monotonically with increasing $E_d$. Within the experimentally relevant range of $\sim$37--48\,eV, this corresponds to a $\sim$10--20\% uncertainty in the number of resulting vacancies. Assuming a fixed vacancy-to-NV conversion efficiency, this rescales the KMC-predicted NV yield from $\sim$23 to $\sim$21 at the upper end of $E_d$, and therefore does not fully account for the experimentally measured value of $\sim$17 NV centers.

\begin{figure}[htbp]
    \centering
    \includegraphics[width=\textwidth]{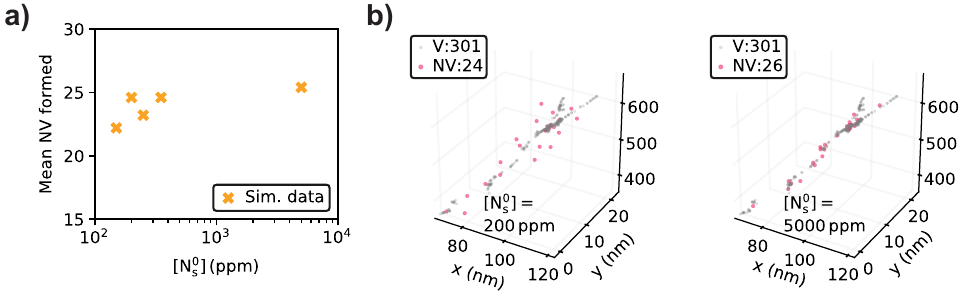}
    \caption{Impact of nitrogen concentration on annealing simulations. (a) Mean NV yield as a function of nitrogen concentration $[\rm{N}_{\rm{s}}^0]$, showing weak dependence over a broad range (150–5000\,\si{ppm}). (b) Representative spatial distribution of vacancies (gray) and resulting NV centers (red) at $[\rm{N}_{\rm{s}}^0]=200\,\rm{ppm}$ and 5000\,ppm. Higher nitrogen concentration leads to NV formation closer to the original damage track, while the overall NV yield remains largely unchanged.} 
    \label{fig:sm_factorsNVyield_2}
\end{figure}

\textbf{Nitrogen concentration.} The measured nitrogen concentration carries a relatively large uncertainty, ranging from $\sim$180--320 ppm as described above. To assess its impact on NV yield, we perform annealing simulations across a range of nitrogen concentrations, as shown in Fig. \ref{fig:sm_factorsNVyield_2}a. For each concentration, a single damage track is simulated, and the nitrogen lattice configuration is randomized five times to estimate the mean NV yield. We find that the mean NV yield changes only weakly over the range from 150\,ppm to 5000\,ppm. Increasing nitrogen concentration primarily leads to NV formation occurring closer to the original damage track, as illustrated in Fig. \ref{fig:sm_factorsNVyield_2}b. This weak dependence can be understood from the spatial scales involved: vacancy and interstitial defects are typically separated by less than $\sim$0.5\,nm, whereas the average nitrogen–nitrogen spacing is $\sim$3\,nm at 200\,ppm and scales with the inverse cubic root of concentration. As a result, vacancies predominantly react with interstitials and other vacancies before diffusing away from the damage track to interact with nitrogen, making the influence of nitrogen concentration on overall NV yield much less pronounced.

\section{DFT calculations of defect interactions}
\label{sec:dft}
\subsection{Methods}

First principles calculations of defect formation in diamond are performed with the Vienna \textit{Ab initio} Simulation Package (VASP) \cite{kresse1993ab,kresse1996efficient}, which implements plane-wave basis sets using a projector-augmented wave (PAW) scheme \cite{blochl1994projector}. We examined two elementary formation mechanisms of vacancy defects in diamond:
\begin{align}
& \mathrm{C}_\mathrm{C} \leftrightarrow \mathrm{V}_\mathrm{C} + \mathrm{C}_\mathrm{i} \quad \text{(Reaction 1)} \\
&\mathrm{V}_{2\mathrm{C}} \leftrightarrow 2\,\mathrm{V}_\mathrm{C} \quad \text{(Reaction 2)}
\end{align}
where Reaction 1 describes the formation of a carbon monovacancy (V$_\mathrm{C}$) and carbon interstitial (C$_\mathrm{i}$) via ejection of a lattice-site carbon atom (C$_\mathrm{C}$), and Reaction 2 describes the ejection of carbon monovacancies from a divacancy site (V$_\mathrm{2C}$).

Defects were embedded within a $3 \times 3 \times 3$ (216 atoms) supercell. Ionic geometries of vacancy defects were relaxed at the Perdew-Burke-Ernzerhof (PBE) level of theory according to the generalized gradient approximation (GGA) \cite{perdew1991generalized,PBE1996}. Sampling of the Brillouin zone for defect structures was done at the $\Gamma$ point with a cutoff energy of 520\,eV. Convergence criteria for electronic self-consistency and ionic relaxation were set to $10^{-10}$\,eV and $10^{-3}$\,eV\,$\text{\AA}^{-1}$, respectively.

Defect formation energies ($E_f$) were calculated according to
\begin{equation}
E_f = E_{\mathrm{def}} - E_{\mathrm{pristine}} + \sum_i n_i \mu_i,
\end{equation}
where $E_{\mathrm{def}}$ and $E_{\mathrm{pristine}}$ denote the energies of the defective and pristine lattices, respectively, and $\mu_i$ and $n_i$ represent the chemical potential and number of atoms corresponding to species $i$. Because the kinetic Monte Carlo (KMC) simulations for annealing do not explicitly account for charge states of defects, only neutral configurations were considered, and no charge correction term is applied.

The energy barriers ($E_B$) associated with Reactions 1 and 2 were determined using a climbing image-modified nudged elastic band (NEB) approach \cite{sheppard2012generalized}. The NEB calculations were carried out using the PBE functional with a spring constant of 5.0\,eV\,$\text{\AA}^{-2}$ between images. Here, $E_f$ and $E_B$ are distinct quantities, where $E_B$ corresponds to the energy difference between the transition state and the initial configuration.

\subsection{Results}

\begin{figure}[htbp]
    \centering
    \includegraphics[width=\textwidth]{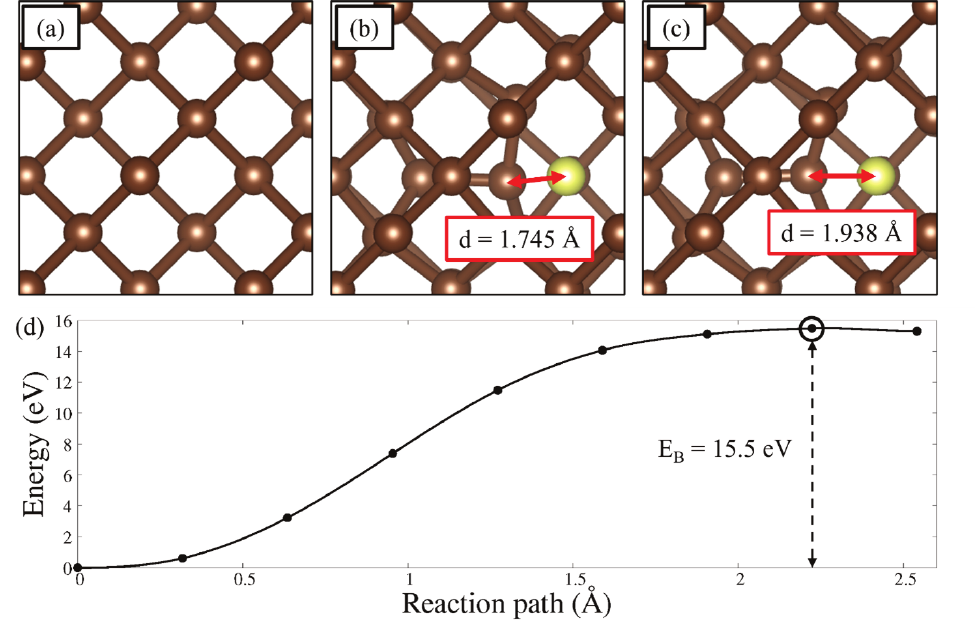}
    \caption{NEB-DFT calculations of Reaction 1. Geometries of the (a) initial, (b) transition geometry, and (c) final states along the reaction path. V$_{\rm{C}}$ and C$_{\rm{i}}$ defects are formed by the ejection of a C$_{\rm{C}}$ atom from its lattice site. Carbon atoms are shown in brown, while the resulting vacancy is shown in yellow. (d) MEP of Reaction 1. The transition state occurs at a V$_{\rm{C}}$-C$_{\rm{i}}$ spacing of $1.75\,\text{\AA}$ corresponding to E$_{\rm{B}}$ = 15.5\,eV.} 
    \label{fig:sm_nebdft_VI}
\end{figure}

\begin{figure}[htbp]
    \centering
    \includegraphics[width=\textwidth]{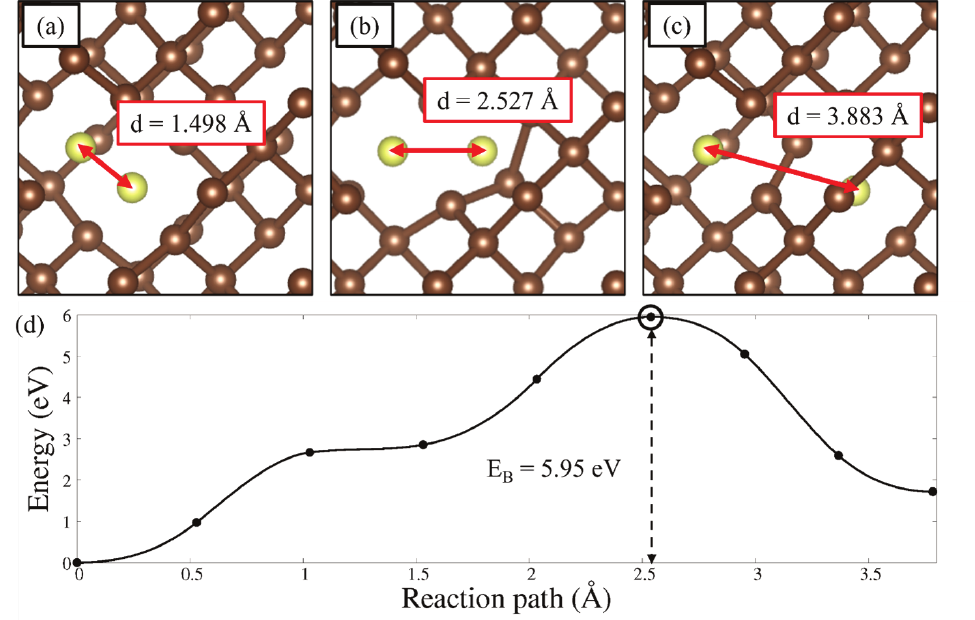}
    \caption{NEB-DFT calculations of Reaction 2. Geometries of the (a) initial, (b) transition geometry, and (c) final states along the reaction path. Two V$_{\rm{C}}$ defects are formed by the dissociation of constituent V$_{\rm{C}}$ from the V$_{\rm{2C}}$ defect. Carbon atoms are shown in brown, while the resulting vacancy is shown in yellow. (d) MEP of Reaction 2. The transition state occurs at a V$_{\rm{C}}$-V$_{\rm{C}}$ spacing of $2.53\,\text{\AA}$ corresponding to E$_{\rm{B}}$ = 5.95\,eV.} 
    \label{fig:sm_nebdft_VV}
\end{figure}

We performed first principles calculations to qualitatively improve the prediction of NV center formation in KMC annealing simulations. Using the NEB method, we elucidate the transition state geometries and associated energy barriers for Reactions 1 and 2. The MEPs corresponding to V$_{\rm{C}}$ formation and V$_{\rm{C}}$ ejection from V$_{\rm{2C}}$ are shown in Figures \ref{fig:sm_nebdft_VI} and \ref{fig:sm_nebdft_VV}, respectively; we have included images of the initial state, transition state, and final state of each reaction pathway, as well.

As shown in Figure \ref{fig:sm_nebdft_VI}, we predict that formation of a V$_{\rm{C}}$ defect and accompanying C$_{\rm{i}}$ defect from a pristine diamond structure is achieved via the formation of a split interstitial located two lattice sites away from the vacancy position. By examining the reaction pathway, we can calculate the radial distance between the ejected  C$_{\rm{i}}$ atom and the resulting  V$_{\rm{C}}$ defect at both the transition state and final state geometries. We have found that at its stable endpoint, the  C$_{\rm{i}}$ defect lies about $1.94\,\text{\AA}$ away from V$_{\rm{C}}$, whereas the  C$_{\rm{i}}$ and V$_{\rm{C}}$ defects are separated by only $1.75\,\text{\AA}$ at the transition state. The transition state is predicted to have a formation energy of approximately 15.5\,eV, and the endpoint has a formation energy of about 15.2\,eV.

The MEP for the dissociation of a V$_{\rm{C}}$ in a V$_{\rm{2C}}$ defect is shown in Figure \ref{fig:sm_nebdft_VV}. Unlike Reaction 1, we find that the constituent V$_{\rm{C}}$ defects require a separation of three lattice sites ($\sim$$3.88\,\text{\AA}$) to stably form two distinct vacancies. Instead, the transition state occurs when the V$_{\rm{C}}$ defects are separated by two lattice sites ($\sim$$2.53\,\text{\AA}$). The dissociation of a V$_{\rm{C}}$ defect from a V$_{\rm{2C}}$ defect is predicted to have an energy barrier of 5.95\,eV, which is in good agreement with previously reported values \cite{slepetz2014divacancies}. 

These NEB results provide a basis for refining the defect–defect interaction distances used in the KMC simulations, which are restricted to first-nearest neighbor (1NN) separation. Vacancy–vacancy interactions can occur spontaneously when defects are within 2NN separation. For vacancy–interstitial pairs, although the transition state occurs at a separation of $1.75\,\text{\AA}$, the relaxed configuration lies only $\sim$0.2\,eV lower in energy. The apparent atom-vacancy separation in this configuration is $1.94\,\text{\AA}$; however, in the KMC lattice picture, the effective interstitial defect position is defined by the arithmetic mean of the two carbon atoms forming the split interstitial, corresponding to a 2NN separation from the vacancy. In the current KMC framework, defect migration barriers are treated as environment-independent, with isolated hopping barriers of 1.5\,eV for interstitials and 2.59\,eV for vacancies. In contrast, the NEB results indicate a barrier of only $\sim$0.2\,eV for V–I recombination from the 2NN configuration. At the annealing temperature of \SI{800}{\celsius}, this reduced barrier leads to a rate enhancement on the order of $\exp([1.5-0.2]\,\rm{eV}/k_BT)\sim10^6$ compared with isolated defect hopping, indicating that 2NN V–I pairs will preferentially approach and recombine.

As a first-order correction to the 1NN assumption in the KMC model, we modify the initial population of free vacancies available for NV formation -- here, “free” denotes vacancies that can diffuse rather than react immediately at the beginning of the annealing simulation. Specifically, in the current model, free vacancies are defined as those without 1NN defect neighbors. In the corrected estimate, we further exclude vacancies that have 2NN vacancy or interstitial neighbors, accounting for enhanced spontaneous divacancy formation and vacancy–interstitial recombination. The resulting reduced vacancy population is then rescaled using the same conversion factor (free-vacancy-to-NV yield) as in the uncorrected model to estimate the NV yield. 

Finally, we note that NEB results indicate that vacancy–nitrogen interactions can also occur at approximately 2NN separation. However, on average, only $\sim$0.2\% of the initial vacancies have a nitrogen partner within 1NN or 2NN distance; therefore, this effect is neglected in the first-order correction.

\section{Machine learning-based directionality analysis}
\subsection{Model and training details}
To extract directional information beyond counting-based metrics, we employ simulation-based inference (SBI) \cite{tejero2020sbi}, a Bayesian framework for inverse problems in which the likelihood $P(X|\theta)$ is not available in closed form. In our case, the inversion maps the defect distribution $X$ to the initial recoil parameters $\theta$ (e.g., energy and head-tail parity). Due to the stochastic nature of the SIIMPL collision cascade and NV formation during annealing, direct evaluation of the likelihood is intractable. Instead, the model learns an approximation to the posterior $P(\theta|X)$. We implement a Mixed Neural Posterior Estimation (MNPE) architecture, which enables joint inference over continuous (energy) and discrete (parity) variables. The model combines a Neural Spline Flow \cite{durkan2019neural} for the continuous parameter with a categorical softmax head for parity classification, implemented using the \texttt{sbi} toolkit \cite{tejero2020sbi}. We note that energy parameters are included when learning the general inversion from simulated damage tracks across different ion implantation energies; however, for predicting head-tail parity of the annealed 800\,keV NV tracks considered here, these parameters are effectively fixed. Parity classification is binary, i.e. the $+z$ or $-z$ direction in the global coordinate system, as the injected carbon ion is along the $z$ axis in the SIIMPL simulation.

Rather than using raw defect coordinates, the model learns from a set of summary features derived from defect spatial distributions. Each track is encoded as a 13-dimensional feature vector (see Table \ref{tab:mnpe_features}) designed to capture directional information. The model is trained on 4000 SIIMPL-simulated vacancy tracks with ground-truth parity labels using the Adam optimizer (learning rate $5\times10^{-4}$, batch size 128) and a 90/10 train-validation split, with early stopping applied when the validation loss does not improve for 20 epochs. For inference, the trained model is directly applied to NV distributions after annealing using the same feature representation. For each track, 200 samples are drawn from $P(\theta|X)$ to estimate the probabilities $P$ of the $+z$ direction, with $1-P$ corresponding to the $-z$ direction. A probability threshold symmetric around 0.5 can then be applied to classify tracks as correct, incorrect, or indeterminate.

\begin{table*}[!tp]
\centering
\caption{The 13 summary features computed for each track and passed to the
MNPE model. Here $N$ is the number of defects and $z_i$ is the position of
the $i$-th defect along the track axis, measured relative to the track's
centroid  (each track is re-centered to $z=0$). $N_\pm$ and $z_\pm$ denote the counts and positions of defects with $z_i > 0$ and $z_i < 0$, respectively.}
\label{tab:mnpe_features}
\renewcommand{\arraystretch}{1.35}
\resizebox{\textwidth}{!}{%
\begin{tabular}{@{}clp{6.5cm}p{6.8cm}@{}}
\toprule
\# & Feature & Definition & Interpretation \\
\midrule
1 & Mean depth
  & $\dfrac{1}{N}\sum_i |z_i|$
  & The average extent of the track. Longer mean depth usually means a higher-energy ion. \\
2 & Max depth
  & $\max_i |z_i|$
  & Distance of the farthest defect from track center. \\
3 & Vacancy count
  & $N$
  & The total number of defects. \\
4\,\textendash\,9 & Depth-profile fractions ($k = 1, \dots, 6$)
  & Fraction of defects in each of six log-spaced bins along the normalized track depth
  & Defect distribution binned along $z$ axis. \\
10 & Centered asymmetry
  & $(N_+ - N_-)/N$
  & Whether more defects sit on one side of the centroid than the other. \\
11 & Asymmetry ratio defined in Ref. \cite{rajendran_method_2017}
  & $N_{\text{last third}} \big/ N_{\text{first third}}$
  & The asymmetry ratio from Ref.~\cite{rajendran_method_2017}. \\
12 & Longitudinal skewness
  & Skewness of $\{z_i\}$ along the track axis
  & Whether the defects lean more toward the head or the tail of the track. \\
13 & Log-variance difference
  & $\log(1 + \mathrm{Var}(z_+)) - \log(1 + \mathrm{Var}(z_-))$
  & Compares how spread out the defects are on each side of the centroid. \\
\bottomrule
\end{tabular}}
\end{table*}

\subsection{Classification performance comparison using NV tracks}
\begin{figure}[htbp]
    \centering
    \includegraphics[width=\textwidth]{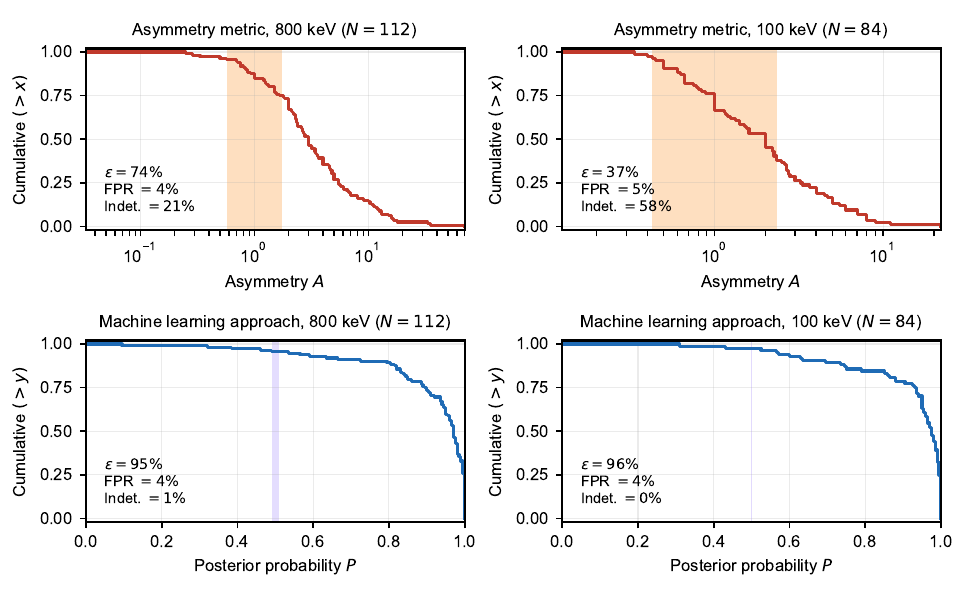}
    \caption{Cumulative plots of NV track classification using the counting-based asymmetry metric and machine learning approach. Top row: cumulative distributions of the asymmetry metric for 800\,keV and 100\,keV NV tracks. Bottom row: cumulative distributions of the machine learning posterior probability for the same tracks. Shaded regions indicate indeterminate ranges chosen to give a $\lesssim$5\% false positive rate. The machine learning approach increases the classification efficiency relative to the asymmetry metric for both implantation energies.} 
    \label{fig:sm_cumulative_all}
\end{figure}

To compare the counting-based asymmetry metric and the machine learning model across implantation energies, we present cumulative plots of NV track classification performance (Fig.~\ref{fig:sm_cumulative_all}). NV tracks are simulated for both 800\,keV and 100\,keV ion implantations. Analogous to Fig.~\ref{fig:siimplkmc}c,d in the main text, thresholds are chosen for the asymmetry metric $A$ and posterior probability $P$ such that the false positive rate is fixed at $\lesssim$5\%. For the machine learning analysis, the model applied to 800\,keV NV tracks is trained on 800\,keV vacancy tracks, while the model applied to 100\,keV NV tracks is separately trained on 100\,keV vacancy tracks. For the 100\,keV case, despite the lower average number of NV centers formed per track, the machine learning approach improves the classification efficiency $\epsilon$ from 37\% to 96\%.

\subsection{Classification performance at low recoil energy}
We further evaluated classification performance at a much lower recoil energy of 1\,keV. In this regime, each track produces only about 11 vacancies on average, and the expected number of NV centers after annealing is $<1$, below the optical detection threshold. Therefore, NV-based readout is not expected to be viable at this energy under the current annealing-based scheme. Nevertheless, vacancy tracks provide a useful benchmark for testing how directional classification degrades at very low recoil energy and for comparing the machine learning (ML) approach with the counting-based asymmetry metric. At a $\lesssim$5\% false positive rate, the machine learning classification efficiency decreases to 26\%, but remains substantially higher than the 6\% efficiency obtained using the asymmetry metric. This suggests that our ML analysis can still extract directional information from sparse damage tracks, and may be useful for future defect candidates that are optically bright, single-defect resolvable, and localizable with high spatial precision without requiring annealing activation.

\begin{figure}[htbp]
    \centering
    \includegraphics[width=\textwidth]{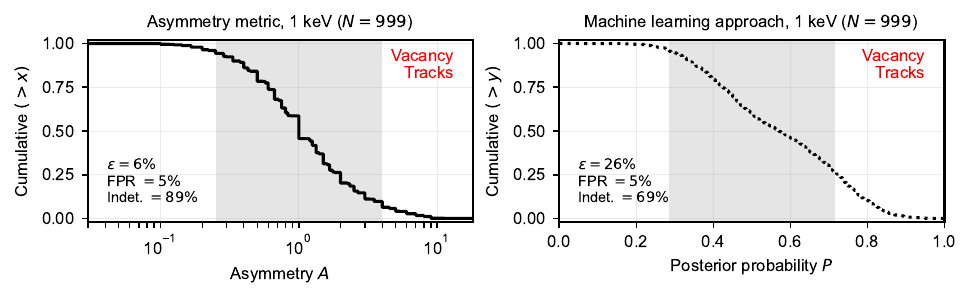}
    \caption{Cumulative plots of vacancy track classification using the counting-based asymmetry metric and machine learning approach. Vacancy tracks are used instead of NV tracks because the expected number of NV centers after annealing is $<1$. At a $\lesssim$5\% false positive rate, the machine learning approach gives higher classification efficiency than the asymmetry metric.} 
    \label{fig:sm_cumulative_1keV}
\end{figure}

\section{Additional spin echo measurement data}
\begin{figure}[htbp]
    \centering
    \includegraphics[width=\textwidth]{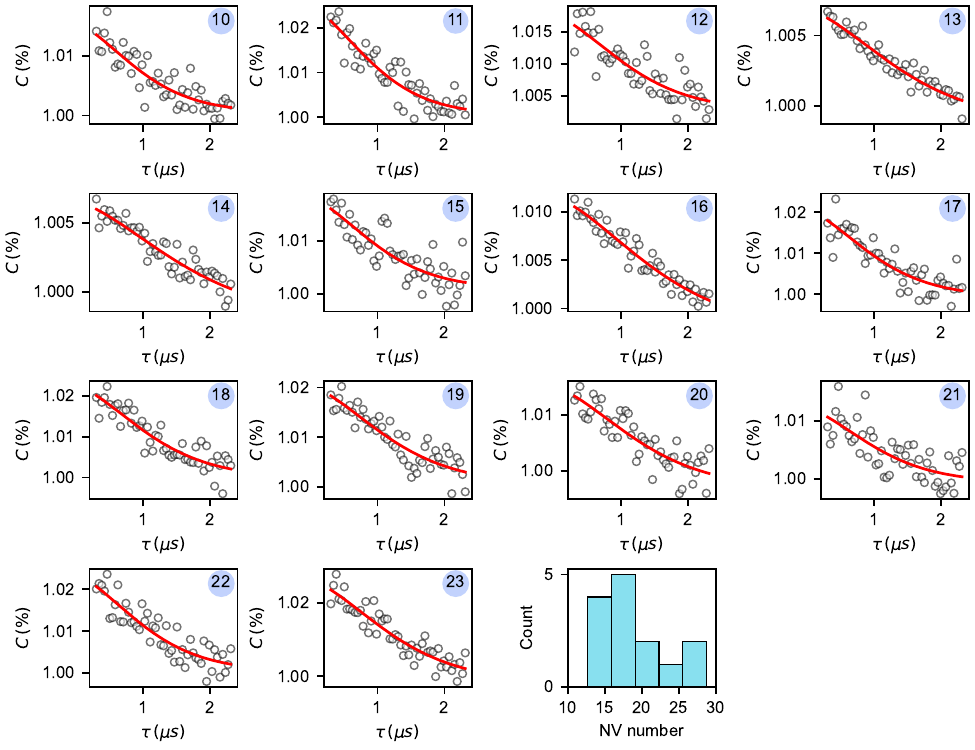}
    \caption{Additional NV spin echo measurements. Spin echo measurements from 14 additional implantation sites in the HPHT sample. Each trace is fitted to a stretched exponential decay (red curve) with fixed exponent $p=1.55$. The final panel shows the distribution of NV PL counts for the characterized sites.} 
    \label{fig:sm_additionalspin_1}
\end{figure}

Additional spin echo measurements are performed on 14 implantation sites in the implanted HPHT sample to investigate potential correlations among spin contrast, coherence properties, and the number of NV centers per site. Individual decay traces and the corresponding NV number distribution are shown in Fig. \ref{fig:sm_additionalspin_1}. These data were acquired with reduced averaging compared to the main text and were fitted using a stretched exponential model with the exponent fixed at $p=1.55$, informed by the main text results, i.e. $C_0 \exp[-(\tau/T_2)^{1.55}]$.

\begin{figure}[htbp]
    \centering
    \includegraphics[width=\textwidth]{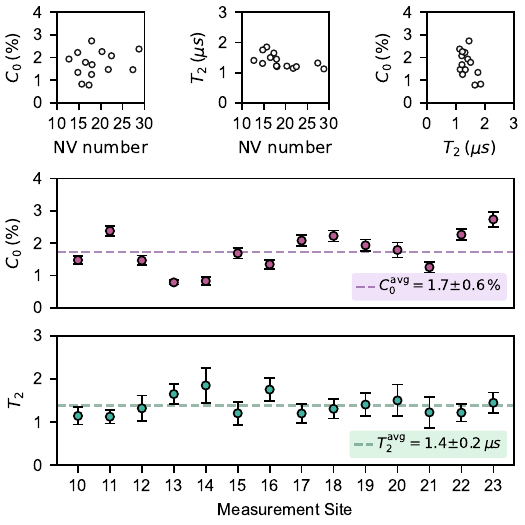}
    \caption{Spin properties versus NV number. (a) Scatter plots of spin-state-dependent PL contrast $C_{0}$, coherence time $T_2$, and NV number per site, showing no clear correlation between spin properties and NV count. (b) Distribution of $C_{0}$ and $T_2$ across measurement sites. $T_2$ remains relatively constant around $1.4\,\mu\mathrm{s}$, while $C_{0}$ exhibits a larger spread ($\sim$1-3\%).}
    \label{fig:sm_additionalspin_2}
\end{figure}

The extracted spin echo parameters and NV numbers are summarized in Fig. \ref{fig:sm_additionalspin_2}. The coherence time $T_2$ remains relatively constant across measurement sites, with an average value of ${\sim}1.4\,\mu\mathrm{s}$. In contrast, the PL contrast exhibits a larger spread ($\sim$1--3\%). Neither shows a clear dependence on NV count. This variation in contrast may arise from differences in the distribution of NV orientations within each site, as the measurement selectively probes a single NV axis while the remaining three crystal orientations contribute to background PL.

\section{Improving diamond surface preparation}
\label{sec:diamondsurfaceprep}

\begin{figure}[htbp]
    \centering
    \includegraphics[width=\textwidth]{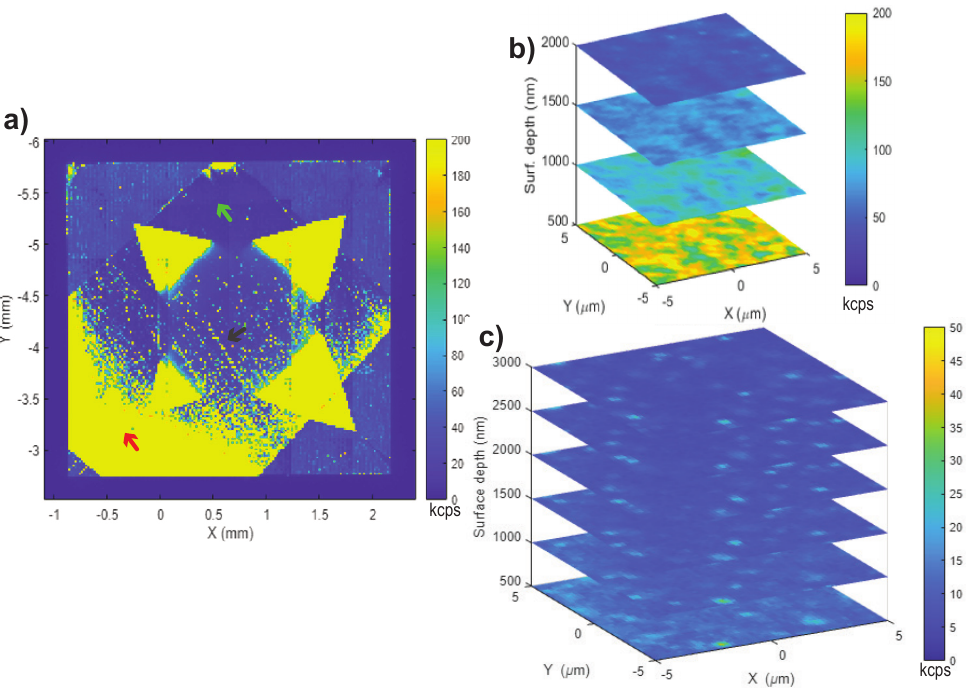}
    \caption{Mitigation of polishing-induced NV background. (a) Confocal PL map of a scaife-polished HPHT diamond measured at a depth of ${\sim}1\,\mu\rm{m}$ after annealing. The diamond is predominantly \{100\}-oriented, with four high-PL “petal” regions arising from growth sectors. Strong and highly non-uniform background PL is observed due to polishing-induced damage. Red arrows indicate regions of severe damage that can overwhelm implantation signals, black arrows highlight streak-like features consisting of isolated high-NV spots, and green arrows indicate regions with relatively low background. Three corners have been previously etched, showing indented features. (b) Confocal PL maps at increasing surface depths following high-power (250\,W) Ar/Cl$_2$ ICP-RIE etching. The near-surface region ($\lesssim1\,\mu\rm{m}$) exhibits elevated PL corresponding to $\sim$5-7 NV centers per confocal spot, indicating etch-induced damage. (c) Confocal PL maps at increasing depths following low-power (20\,W) Ar/Cl$_2$ ICP-RIE etching. The near-surface PL is significantly reduced, with NV densities $\sim$1 per confocal spot, comparable to as-received HPHT diamond without annealing.} 
    \label{fig:sm_surfaceprep}
\end{figure}

The NV count analysis of Section~\ref{si:nv} required manually excluding sub-regions with high NV backgrounds from residual subsurface polishing damage. These backgrounds arise from the diamond polishing process. Synthetic HPHT diamond is typically laser-cut into individual plates and mechanically polished to achieve optically smooth surfaces. However, mechanical polishing is known to introduce subsurface damage \cite{schuelke2013diamond}, generating lattice vacancies that can form NV centers after annealing. In our ion implantation studies, where damage tracks are formed near the surface, such polishing-induced defects can obscure or interfere with the implantation signal. As shown in Fig. \ref{fig:sm_surfaceprep}a, a scaife-polished diamond surface (with roughness $R_a\lesssim1\,\rm{nm}$) exhibits strong and highly non-uniform photoluminescence after annealing. Although regions with low background can still be found after annealing, their spatial distribution is difficult to control. In contrast, prior to annealing, the same material typically exhibits fewer than $\sim$2 NV centers per confocal spot across the sample surface, except for the non-\{100\} sectors (e.g., the four "petal" features near the diamond center). 

We performed several experiments to suppress this background through improved surface preparation. These procedures were developed after the implanted sample measured in this work and are intended for future studies. The primary approach we investigated was argon–chlorine (Ar/Cl$_2$) inductively coupled plasma reactive ion etching (ICP-RIE) \cite{appel2016fabrication, zhou2017scanning,lee2008etching}, which removes polishing-induced damage from the near-surface region. While such etching processes have shown success in low-nitrogen CVD diamond, their application to high-nitrogen HPHT diamond ($[\rm{N}_{\rm{s}}^0]\sim200\,\rm{ppm}$) requires further evaluation, as etching-induced damage may be more pronounced. We performed etching with varying RF power (Table \ref{tab:etch_recipes}), followed by annealing and PL imaging at different depths to quantify the resulting NV density. In each case, more than ${\sim}10\,\mu\rm{m}$ of material is removed to eliminate the original polishing-damaged layer. For high RF power (250\,W), the NV density peaks at $\sim$5--7 per confocal spot at a depth of ${\sim}500\,\rm{nm}$ (Fig. \ref{fig:sm_surfaceprep}b), whereas reducing the power to 20\,W lowers the density to $\sim$1 per spot (Fig. \ref{fig:sm_surfaceprep}c), comparable to the as-received HPHT diamond prior to annealing. Further reduction in RF power yields diminishing improvement while reducing the etch rate. We note that similar surface NV density can also be achieved with shorter etching time by combining an initial high-power etch with a subsequent low-power step to remove 3--4$\,\mu\rm{m}$ of fast-etch-induced damage.

\begin{table}[htbp]\centering\caption{Ar/Cl$_2$ ICP-RIE etching recipes and O$_2$ surface finishing step.}\begin{tabularx}{\linewidth}{LccC{1.3cm}C{1.4cm}C{1.3cm}C{1.5cm}}\hline\textbf{Step} &\textbf{Gas} &\makecell{\textbf{Flow}\\\textbf{(sccm)}} &\makecell{\textbf{RF}\\\textbf{Power}\\\textbf{(W)}} &\makecell{\textbf{ICP}\\\textbf{Power}\\\textbf{(W)}} &\makecell{\textbf{Etch}\\\textbf{Rate}\\\textbf{(nm/hr)}} &\makecell{\textbf{NVs /}\\\textbf{confocal}\\\textbf{spot}} \\\hline Etch Recipe 1 & Ar/Cl$_2$ & 25 / 40 & 250 & 400 & 3000 & $\sim$5-7 \\Etch Recipe 2 & Ar/Cl$_2$ & 25 / 40 & 20  & 400 & 480  & $\lesssim1$ \\Etch Recipe 3 & Ar/Cl$_2$ & 25 / 40 & 8   & 400 & 117  & $\lesssim1$ \\\hline O$_2$ finish & O$_2$ & 30 & 20 & 700 & -- & -- \\\hline\end{tabularx}\label{tab:etch_recipes}\end{table}

\clearpage

\bibliography{darkmatterbiblio}

@article{tagami2005,
  author    = {Tagami, Takahiro and O'Sullivan, Paul B.},
  title     = {Fundamentals of Fission-Track Thermochronology},
  journal   = {Reviews in Mineralogy and Geochemistry},
  volume    = {58},
  number    = {1},
  pages     = {19--47},
  year      = {2005},
  doi       = {10.2138/rmg.2005.58.2},
}

@article{Delegan_2023,
doi = {10.1088/1361-6528/acdd09},
url = {https://doi.org/10.1088/1361-6528/acdd09},
year = {2023},
month = {jul},
publisher = {IOP Publishing},
volume = {34},
number = {38},
pages = {385001},
author = {Delegan, Nazar and Whiteley, Samuel J and Zhou, Tao and Bayliss, Sam L and Titze, Michael and Bielejec, Edward and Holt, Martin V and Awschalom, David D and Heremans, F Joseph},
title = {Deterministic nanoscale quantum spin-defect implantation and diffraction strain imaging},
journal = {Nanotechnology}
}

@article{Karger_2018,
doi = {10.1088/1361-6560/aa9102},
url = {https://doi.org/10.1088/1361-6560/aa9102},
year = {2017},
month = {dec},
publisher = {IOP Publishing},
volume = {63},
number = {1},
pages = {01TR02},
author = {Karger, Christian P and Peschke, Peter},
title = {{RBE} and related modeling in carbon-ion therapy},
journal = {Physics in Medicine \& Biology}
}

@techreport{fastenau1982diffusion,
	title        = {Diffusion limited reactions in crystalline solids},
	author       = {Fastenau, R and others},
	year         = 1982,
	institution  = {Technische Hogeschool Delft (Netherlands)}
}

@article{koike1992displacement,
	title        = {Displacement threshold energy for type IIa diamond},
	author       = {Koike, Junichi and Parkin, DM and Mitchell, TE},
	year         = 1992,
	journal      = {Applied Physics Letters},
	publisher    = {American Institute of Physics},
	volume       = 60,
	number       = 12,
	pages        = {1450--1452}
}

@article{slepetz2014divacancies,
	title        = {Divacancies in diamond: a stepwise formation mechanism},
	author       = {Slepetz, Brad and Kertesz, Miklos},
	year         = 2014,
	journal      = {Physical Chemistry Chemical Physics},
	publisher    = {Royal Society of Chemistry},
	volume       = 16,
	number       = 4,
	pages        = {1515--1521}
}

@article{wierbik2026anisotropic,
	title = {Anisotropic fine structure of ion tracks in single crystals},
	volume = {113},
	url = {https://link.aps.org/doi/10.1103/66z7-gjzq},
	doi = {10.1103/66z7-gjzq},
	number = {3},
	urldate = {2026-06-03},
	journal = {Physical Review B},
	publisher = {American Physical Society},
	author = {Wierbik, Jessica and Heimes, Hendrik and Notthoff, Christian and Dutt, Shankar and Alwadi, Taleb and Kiy, Alexander and Mota-Santiago, Pablo and Kirby, Nigel and Kluth, Patrick},
	month = jan,
	year = {2026},
	pages = {035306},
}

@article{kim2025athermal,
	title        = {Athermal phonon collection efficiency in diamond crystals for low mass dark matter detection},
	author       = {Kim, I and Kurinsky, NA and Kagan, H and Boyd, STP and Kim, GB},
	year         = 2025,
	journal      = {Physical Review D},
	publisher    = {APS},
	volume       = 111,
	number       = 7,
	pages        = {072009}
}

@article{kim2025scalable,
	title        = {Scalable nanoscale positioning of highly coherent color centers in prefabricated diamond nanostructures},
	author       = {Kim, Sunghoon and London, Paz and Yang, Daipeng and Hughes, Lillian B and Ahlers, Jeffrey and Meynell, Simon and Mitchell, William J and Mukherjee, Kunal and Bleszynski Jayich, Ania C},
	year         = 2025,
	journal      = {Nature Communications},
	publisher    = {Nature Publishing Group UK London},
	volume       = 16,
	number       = 1,
	pages        = 9803
}

@article{vanEnckevort1988,
	title        = {Photoluminescence tomography as a method to image point-defect distributions in crystals: Nitrogen-vacancy pairs in synthetic diamonds},
	author       = {van Enckevort, W. J. P. and Lochs, H. G. M.},
	year         = 1988,
	journal      = {Journal of Applied Physics},
	volume       = 64,
	number       = 1,
	pages        = {434--437},
	doi          = {10.1063/1.341212}
}

@article{luhmann2019coulomb,
	title        = {Coulomb-driven single defect engineering for scalable qubits and spin sensors in diamond},
	author       = {L{\"u}hmann, Tobias and John, Roger and Wunderlich, Ralf and Meijer, Jan and Pezzagna, S{\'e}bastien},
	year         = 2019,
	journal      = {Nature Communications},
	publisher    = {Nature Publishing Group UK London},
	volume       = 10,
	number       = 1,
	pages        = 4956
}

@article{lang2020fundamental,
	title        = {Fundamental Phenomena and Applications of Swift Heavy Ion Irradiations},
	author       = {Lang, Maik and Djurabekova, Flyura and Medvedev, Nikita and Toulemonde, Marcel and Trautmann, Christina},
	year         = 2020,
	month        = {08},
	journal      = {Comprehensive Nuclear Materials},
	volume       = 1,
	doi          = {10.1016/b978-0-12-803581-8.11644-3},
	url          = {https://www.osti.gov/biblio/1846448},
	place        = {United States}
}

@article{mitchell2023spparks,
	title        = {Parallel simulation via {SPPARKS} of on‑lattice kinetic and {Metropolis Monte Carlo} models for materials processing},
	author       = {Mitchell, John A. and Abdeljawad, Fadi and Battaile, Corbett and Garcia-Cardona, Cristina and Holm, Elizabeth A. and Homer, Eric R. and Madison, Jon and Rodgers, Theron M. and Thompson, Aidan P. and Tikare, Veena and Webb, Ed and Plimpton, Steven J.},
	year         = 2023,
	journal      = {Modelling and Simulation in Materials Science and Engineering},
	volume       = 31,
	number       = 5,
	pages        = {055001},
	doi          = {10.1088/1361-651X/accc4b},
	url          = {https://doi.org/10.1088/1361-651X/accc4b}
}

@phdthesis{janson2003hydrogen,
	title        = {Hydrogen diffusion and ion implantation in silicon carbide},
	author       = {Janson, M. S.},
	year         = 2003,
	issn         = {0284-0545},
	url          = {http://www.diva-portal.org/smash/get/diva2:9286/FULLTEXT01.pdf},
	school       = {KTH-Royal Institute of Technology},
	type         = {{Ph.D. thesis}}
}

@misc{siimpl,
	title        = {{SIIMPL: Simulation of Ion IMPLantation}},
	author       = {Janson, M. S.},
	year         = 2022,
	note         = {Accessed: 2025-07-23},
	howpublished = {\url{https://github.com/msjanson01/siimpl}}
}

@article{yamamotoIsotopicIdentificationEngineered2014,
	title        = {Isotopic identification of engineered nitrogen-vacancy spin qubits in ultrapure diamond},
	author       = {Yamamoto, T. and Onoda, S. and Ohshima, T. and Teraji, T. and Watanabe, K. and Koizumi, S. and Umeda, T. and McGuinness, L. P. and Müller, C. and Naydenov, B. and Dolde, F. and Fedder, H. and Honert, J. and Markham, M. L. and Twitchen, D. J. and Wrachtrup, J. and Jelezko, F. and Isoya, J.},
	year         = 2014,
	month        = aug,
	journal      = {Physical Review B},
	volume       = 90,
	number       = 8,
	pages        = {081117},
	doi          = {10.1103/PhysRevB.90.081117},
	url          = {https://link.aps.org/doi/10.1103/PhysRevB.90.081117},
	urldate      = {2025-02-26}
}

@article{rackeVacancyDiffusionNitrogenvacancy2021,
	title        = {Vacancy diffusion and nitrogen-vacancy center formation near the diamond surface},
	author       = {Räcke, P. and Pietzonka, L. and Meijer, J. and Spemann, D. and Wunderlich, R.},
	year         = 2021,
	month        = may,
	journal      = {Applied Physics Letters},
	volume       = 118,
	number       = 20,
	pages        = 204003,
	doi          = {10.1063/5.0046031},
	issn         = {0003-6951},
	url          = {https://doi.org/10.1063/5.0046031},
	urldate      = {2025-02-26}
}

@article{deakFormationNVCenters2014,
	title        = {Formation of {NV} centers in diamond: {A} theoretical study based on calculated transitions and migration of nitrogen and vacancy related defects},
	shorttitle   = {Formation of {NV} centers in diamond},
	author       = {Deák, Peter and Aradi, Bálint and Kaviani, Moloud and Frauenheim, Thomas and Gali, Adam},
	year         = 2014,
	month        = feb,
	journal      = {Physical Review B},
	volume       = 89,
	number       = 7,
	pages        = {075203},
	doi          = {10.1103/PhysRevB.89.075203},
	url          = {https://link.aps.org/doi/10.1103/PhysRevB.89.075203},
	urldate      = {2025-10-17}
}

@misc{araujoNuclearRecoilDetection2025,
	title        = {Nuclear recoil detection with color centers in bulk lithium fluoride},
	author       = {Araujo, Gabriela A. and Baudis, Laura and Bowden, Nathaniel and Chapman, Jordan and Erickson, Anna and Perez, Mariano Guerrero and Hecht, Adam A. and Hedges, Samuel C. and Huber, Patrick and Ivanov, Vsevolod and Jovanovic, Igor and Khodaparast, Giti A. and Magill, Brenden A. and Mateos, Jose Maria and Morrison, Maverick and Smith, Nicholas W. G. and Stengel, Patrick and Surani, Stuti and Vladimirov, Nikita and Walkup, Keegan and Wittweg, Christian and Zhang, Xianyi},
	year         = 2025,
	month        = mar,
	publisher    = {arXiv},
	doi          = {10.48550/arXiv.2503.20732},
	url          = {http://arxiv.org/abs/2503.20732},
	urldate      = {2025-10-17},
	note         = {arXiv:2503.20732 [nucl-ex]}
}

@article{favarodeoliveiraTailoringSpinDefects2017,
	title        = {Tailoring spin defects in diamond by lattice charging},
	author       = {Fávaro de Oliveira, Felipe and Antonov, Denis and Wang, Ya and Neumann, Philipp and Momenzadeh, Seyed Ali and Häußermann, Timo and Pasquarelli, Alberto and Denisenko, Andrej and Wrachtrup, Jörg},
	year         = 2017,
	month        = may,
	journal      = {Nature Communications},
	volume       = 8,
	number       = 1,
	pages        = 15409,
	doi          = {10.1038/ncomms15409},
	issn         = {2041-1723},
	url          = {https://www.nature.com/articles/ncomms15409},
	urldate      = {2025-10-16},
	copyright    = {2017 The Author(s)},
	language     = {en}
}

@article{strigaricevnsdirectional2020,
	title        = {Coherent elastic neutrino-nucleus scattering with directional detectors},
	author       = {Abdullah, M. and Aristizabal Sierra, D. and Dutta, Bhaskar and Strigari, Louis E.},
	year         = 2020,
	month        = jul,
	journal      = {Physical Review D},
	volume       = 102,
	number       = 1,
	pages        = {015009},
	doi          = {10.1103/PhysRevD.102.015009},
	url          = {https://link.aps.org/doi/10.1103/PhysRevD.102.015009},
	urldate      = {2025-11-11}
}

@article{MARTIN1981,
	title        = {Electrical rotation of quadrupole lenses},
	author       = {F.W. Martin},
	year         = 1981,
	journal      = {Nuclear Instruments and Methods in Physics Research},
	volume       = 189,
	number       = 1,
	pages        = {93--96},
	doi          = {https://doi.org/10.1016/0029-554X(81)90128-2},
	issn         = {0167-5087},
	url          = {https://www.sciencedirect.com/science/article/pii/0029554X81901282}
}

@article{Chu:2006,
	title        = {Ion-microbeam probe of high-speed shift registers for {SEE} Analysis-part I: SiGe},
	author       = {Chu, P. and Hansen, D.L. and Doyle, B.L. and Jobe, K. and Lopez-Aguado, R. and Shoga, M. and Walsh, D.S.},
	year         = 2006,
	journal      = {IEEE Transactions on Nuclear Science},
	volume       = 53,
	number       = 3,
	pages        = {1574--1582},
	doi          = {10.1109/TNS.2005.861420}
}

@inproceedings{Ang2024,
	title        = {Progress Toward a Solid-State Directional Dark Matter Detector},
	author       = {Ang, Daniel G. and Liu, Xingxin and Tang, Jiashen and Shen, Maximilian and Ebadi, Reza and Walsworth, Ronald},
	year         = 2024,
	booktitle    = {MDvDM 2024 Proceedings},
	pages        = {18--21},
	doi          = {10.48550/arXiv.2405.01626},
	url          = {https://arxiv.org/abs/2405.01626},
	note         = {arXiv:2405.01626}
}

@inproceedings{Ang2025,
	title        = {Progress Toward a Solid-State Directional Dark Matter Detector},
	author       = {Ang, Daniel G. and Tang, Jiashen and Shen, Maximilian and Camp, Mason and Gilpin, Andrew and Liyanage, Gavishta and Walsworth, Ronald},
	year         = 2025,
	booktitle    = {MDvDM 2025 Proceedings},
	pages        = {18--21},
	doi          = {10.48550/arXiv.2508.20482},
    url          = {https://arxiv.org/abs/2508.20482},
    note         = {arXiv:2508.20482}
}

@article{pezzagna_creation_2010,
	title        = {Creation efficiency of nitrogen-vacancy centres in diamond},
	author       = {Pezzagna, S and Naydenov, B and Jelezko, F and Wrachtrup, J and Meijer, J},
	year         = 2010,
	month        = jun,
	journal      = {New Journal of Physics},
	volume       = 12,
	number       = 6,
	pages        = {065017},
	doi          = {10.1088/1367-2630/12/6/065017},
	issn         = {1367-2630},
	url          = {https://doi.org/10.1088/1367-2630/12/6/065017},
	urldate      = {2025-10-30},
	language     = {en}
}

@article{PBE1996,
  title = {Generalized Gradient Approximation Made Simple},
  author = {Perdew, John P. and Burke, Kieron and Ernzerhof, Matthias},
  journal = {Physical Review Letters},
  volume = {77},
  issue = {18},
  pages = {3865--3868},
  numpages = {0},
  year = {1996},
  month = {Oct},
  publisher = {American Physical Society},
  doi = {10.1103/PhysRevLett.77.3865},
  url = {https://link.aps.org/doi/10.1103/PhysRevLett.77.3865}
}

@article{liu_optical_2025,
	title = {Optical and spin properties of nitrogen vacancy centers in diamond formed along high-energy heavy ion tracks},
	volume = {6},
	copyright = {2025 This is a U.S. Government work and not under copyright protection in the US; foreign copyright protection may apply},
	issn = {2662-4443},
	url = {https://www.nature.com/articles/s43246-025-00961-6},
	doi = {10.1038/s43246-025-00961-6},
	language = {en},
	number = {1},
	urldate = {2026-06-22},
	journal = {Communications Materials},
	author = {Liu, Wei and Leino, Aleksi A. M. and Persaud, Arun and Ji, Qing and Jhuria, Kaushalya and Barnard, Edward S. and Aloni, Shaul and Trautmann, Christina and Tomut, Marilena and Wunderlich, Ralf and Nozais, Chloé and Mogan, Saahit and Ocker, Hunter and Anand, Nishanth and Hao, Zhao and Djurabekova, Flyura and Schenkel, Thomas},
	month = nov,
	year = {2025},
	note = {Publisher: Nature Publishing Group},
	pages = {242},
}

@article{lake_direct_2021,
	title        = {Direct formation of nitrogen-vacancy centers in nitrogen doped diamond along the trajectories of swift heavy ions},
	author       = {Lake, Russell E. and Persaud, Arun and Christian, Casey and Barnard, Edward S. and Chan, Emory M. and Bettiol, Andrew A. and Tomut, Marilena and Trautmann, Christina and Schenkel, Thomas},
	year         = 2021,
	month        = feb,
	journal      = {Applied Physics Letters},
	volume       = 118,
	number       = 8,
	pages        = {084002},
	doi          = {10.1063/5.0036643},
	issn         = {0003-6951},
	url          = {https://doi.org/10.1063/5.0036643},
	urldate      = {2025-10-30}
}

@article{nordlund_channeling_2016,
	title        = {Large fraction of crystal directions leads to ion channeling},
	author       = {Nordlund, K. and Djurabekova, F. and Hobler, G.},
	year         = 2016,
	month        = dec,
	journal      = {Physical Review B},
	volume       = 94,
	number       = 21,
	pages        = 214109,
	doi          = {10.1103/PhysRevB.94.214109},
	url          = {https://link.aps.org/doi/10.1103/PhysRevB.94.214109},
	urldate      = {2025-01-03}
}

@article{amekura_latent_2024,
	title        = {Latent ion tracks were finally observed in diamond},
	author       = {Amekura, H. and Chettah, A. and Narumi, K. and Chiba, A. and Hirano, Y. and Yamada, K. and Yamamoto, S. and Leino, A. A. and Djurabekova, F. and Nordlund, K. and Ishikawa, N. and Okubo, N. and Saitoh, Y.},
	year         = 2024,
	month        = feb,
	journal      = {Nature Communications},
	volume       = 15,
	number       = 1,
	pages        = 1786,
	doi          = {10.1038/s41467-024-45934-4},
	issn         = {2041-1723},
	url          = {https://www.nature.com/articles/s41467-024-45934-4},
	urldate      = {2025-10-30},
	copyright    = {2024 The Author(s)},
	language     = {en}
}

@article{ZieglerSRIM2010,
	title        = {{SRIM} – The stopping and range of ions in matter (2010)},
	author       = {James F. Ziegler and M.D. Ziegler and J.P. Biersack},
	year         = 2010,
	journal      = {Nuclear Instruments and Methods in Physics Research Section B: Beam Interactions with Materials and Atoms},
	volume       = 268,
	number       = 11,
	pages        = {1818--1823},
	doi          = {https://doi.org/10.1016/j.nimb.2010.02.091},
	issn         = {0168-583X},
	url          = {https://www.sciencedirect.com/science/article/pii/S0168583X10001862},
	note         = {19th International Conference on Ion Beam Analysis}
}

@article{vahsenDirectionalRecoilDetection2021,
	title        = {Directional {Recoil} {Detection}},
	author       = {Vahsen, Sven E. and O'Hare, Ciaran A.J. and Loomba, Dinesh},
	year         = 2021,
	journal      = {Annual Review of Nuclear and Particle Science},
	volume       = 71,
	number       = 1,
	pages        = {189--224},
	doi          = {10.1146/annurev-nucl-020821-035016},
	url          = {https://doi.org/10.1146/annurev-nucl-020821-035016},
	urldate      = {2024-02-26},
	note         = {\_eprint: https://doi.org/10.1146/annurev-nucl-020821-035016}
}

@article{BarrySensOpt2020,
	title        = {Sensitivity optimization for {NV}-diamond magnetometry},
	author       = {Barry, John F. and Schloss, Jennifer M. and Bauch, Erik and Turner, Matthew J. and Hart, Connor A. and Pham, Linh M. and Walsworth, Ronald L.},
	year         = 2020,
	month        = {Mar},
	journal      = {Reviews of Modern Physics},
	publisher    = {American Physical Society},
	volume       = 92,
	pages        = {015004},
	doi          = {10.1103/RevModPhys.92.015004},
	issue        = 1,
	numpages     = 68
}

@article{arai_fourier_2015,
	title        = {Fourier magnetic imaging with nanoscale resolution and compressed sensing speed-up using electronic spins in diamond},
	author       = {Arai, K. and Belthangady, C. and Zhang, H. and Bar-Gill, N. and DeVience, S. J. and Cappellaro, P. and Yacoby, A. and Walsworth, R. L.},
	year         = 2015,
	month        = oct,
	journal      = {Nature Nanotechnology},
	volume       = 10,
	number       = 10,
	pages        = {859--864},
	doi          = {10.1038/nnano.2015.171},
	issn         = {1748-3395},
	url          = {https://www.nature.com/articles/nnano.2015.171},
	urldate      = {2024-02-26},
	copyright    = {2015 Springer Nature Limited},
	language     = {en}
}

@article{zhang_selective_2017,
	title        = {Selective addressing of solid-state spins at the nanoscale via magnetic resonance frequency encoding},
	author       = {Zhang, H. and Arai, K. and Belthangady, C. and Jaskula, J.-C. and Walsworth, R. L.},
	year         = 2017,
	month        = aug,
	journal      = {npj Quantum Information},
	volume       = 3,
	number       = 1,
	pages        = {1--8},
	doi          = {10.1038/s41534-017-0033-3},
	issn         = {2056-6387},
	url          = {https://www.nature.com/articles/s41534-017-0033-3},
	urldate      = {2024-02-26},
	copyright    = {2017 The Author(s)},
	language     = {en}
}

@article{kresse1993ab,
	title        = {Ab initio molecular dynamics for liquid metals},
	author       = {Kresse, Georg and Hafner, J{\"u}rgen},
	year         = 1993,
	journal      = {Physical Review B},
	publisher    = {APS},
	volume       = 47,
	number       = 1,
	pages        = 558
}

@article{blochl1994projector,
	title        = {Projector augmented-wave method},
	author       = {Bl{\"o}chl, Peter E},
	year         = 1994,
	journal      = {Physical Review B},
	publisher    = {APS},
	volume       = 50,
	number       = 24,
	pages        = 17953
}

@article{perdew1991generalized,
	title        = {Generalized gradient approximations for exchange and correlation: A look backward and forward},
	author       = {Perdew, John P},
	year         = 1991,
	journal      = {Physica B: Condensed Matter},
	publisher    = {Elsevier},
	volume       = 172,
	number       = {1-2},
	pages        = {1--6}
}

@article{sheppard2012generalized,
	title        = {A generalized solid-state nudged elastic band method},
	author       = {Sheppard, Daniel and Xiao, Penghao and Chemelewski, William and Johnson, Duane D and Henkelman, Graeme},
	year         = 2012,
	journal      = {The Journal of Chemical Physics},
	publisher    = {AIP Publishing},
	volume       = 136,
	number       = 7
}

@article{kresse1996efficient,
	title        = {Efficient iterative schemes for ab initio total-energy calculations using a plane-wave basis set},
	author       = {Kresse, Georg and Furthm{\"u}ller, J{\"u}rgen},
	year         = 1996,
	journal      = {Physical Review B},
	publisher    = {APS},
	volume       = 54,
	number       = 16,
	pages        = 11169
}

@article{schuelke2013diamond,
	title        = {Diamond polishing},
	author       = {Schuelke, Thomas and Grotjohn, Timothy A},
	year         = 2013,
	journal      = {Diamond and Related Materials},
	publisher    = {Elsevier},
	volume       = 32,
	pages        = {17--26}
}

@article{durkan2019neural,
	title        = {Neural spline flows},
	author       = {Durkan, Conor and Bekasov, Artur and Murray, Iain and Papamakarios, George},
	year         = 2019,
	journal      = {Advances in Neural Information Processing Systems},
	volume       = 32
}

@article{lee2008etching,
	title        = {Etching and micro-optics fabrication in diamond using chlorine-based inductively-coupled plasma},
	author       = {Lee, CL and Gu, E and Dawson, MD and Friel, I and Scarsbrook, GA},
	year         = 2008,
	journal      = {Diamond and Related Materials},
	publisher    = {Elsevier},
	volume       = 17,
	number       = {7-10},
	pages        = {1292--1296}
}

@article{zhou2017scanning,
	title        = {Scanning diamond {NV} center probes compatible with conventional {AFM} technology},
	author       = {Zhou, Tony X and St{\"o}hr, Rainer J and Yacoby, Amir},
	year         = 2017,
	journal      = {Applied Physics Letters},
	publisher    = {AIP Publishing},
	volume       = 111,
	number       = 16
}

@article{appel2016fabrication,
	title        = {Fabrication of all diamond scanning probes for nanoscale magnetometry},
	author       = {Appel, Patrick and Neu, Elke and Ganzhorn, Marc and Barfuss, Arne and Batzer, Marietta and Gratz, Micha and Tsch{\"o}pe, Andreas and Maletinsky, Patrick},
	year         = 2016,
	journal      = {Review of Scientific Instruments},
	publisher    = {AIP Publishing},
	volume       = 87,
	number       = 6
}

@misc{tejero2020sbi,
	title = {{SBI} -- {A} toolkit for simulation-based inference},
	url = {http://arxiv.org/abs/2007.09114},
	doi = {10.48550/arXiv.2007.09114},
	urldate = {2026-06-22},
	publisher = {arXiv},
	author = {Tejero-Cantero, Alvaro and Boelts, Jan and Deistler, Michael and Lueckmann, Jan-Matthis and Durkan, Conor and Gonçalves, Pedro J. and Greenberg, David S. and Macke, Jakob H.},
	month = jul,
	year = {2020},
	note = {arXiv:2007.09114 [cs.LG]},
}

@article{breuer1995ab,
	title        = {Ab initio investigation of the native defects in diamond and self-diffusion},
	author       = {Breuer, SJ and Briddon, PR},
	year         = 1995,
	journal      = {Physical Review B},
	publisher    = {APS},
	volume       = 51,
	number       = 11,
	pages        = 6984
}

@article{fleischer1965solid,
	title        = {Solid-state track detectors: applications to nuclear science and geophysics},
	author       = {Fleischer, RL and Price, PB and Walker, RM},
	year         = 1965,
	journal      = {Annual Review of Nuclear Science},
	publisher    = {Annual Reviews 4139 El Camino Way, PO Box 10139, Palo Alto, CA 94303-0139, USA},
	volume       = 15,
	number       = 1,
	pages        = {1--28}
}

@article{bauch2018ultralong,
	title        = {Ultralong dephasing times in solid-state spin ensembles via quantum control},
	author       = {Bauch, Erik and Hart, Connor A and Schloss, Jennifer M and Turner, Matthew J and Barry, John F and Kehayias, Pauli and Singh, Swati and Walsworth, Ronald L},
	year         = 2018,
	journal      = {Physical Review X},
	publisher    = {APS},
	volume       = 8,
	number       = 3,
	pages        = {031025}
}

@article{dohertyNitrogenvacancyColourCentre2013,
	title        = {The nitrogen-vacancy colour centre in diamond},
	author       = {Doherty, Marcus W. and Manson, Neil B. and Delaney, Paul and Jelezko, Fedor and Wrachtrup, Jörg and Hollenberg, Lloyd C. L.},
	year         = 2013,
	month        = jul,
	journal      = {Physics Reports},
	series       = {The nitrogen-vacancy colour centre in diamond},
	volume       = 528,
	number       = 1,
	pages        = {1--45},
	doi          = {10.1016/j.physrep.2013.02.001},
	issn         = {0370-1573},
	url          = {https://www.sciencedirect.com/science/article/pii/S0370157313000562},
	urldate      = {2024-02-27}
}

@article{rajendran_method_2017,
	title        = {A method for directional detection of dark matter using spectroscopy of crystal defects},
	author       = {Rajendran, Surjeet and Zobrist, Nicholas and Sushkov, Alexander O. and Walsworth, Ronald and Lukin, Mikhail},
	year         = 2017,
	month        = aug,
	journal      = {Physical Review D},
	volume       = 96,
	number       = 3,
	pages        = {035009},
	doi          = {10.1103/PhysRevD.96.035009},
	urldate      = {2024-01-23}
}

@article{ohareNewDefinitionNeutrino2021,
	title        = {New {Definition} of the {Neutrino} {Floor} for {Direct} {Dark} {Matter} {Searches}},
	author       = {O’Hare, Ciaran A. J.},
	year         = 2021,
	month        = dec,
	journal      = {Physical Review Letters},
	volume       = 127,
	number       = 25,
	pages        = 251802,
	doi          = {10.1103/PhysRevLett.127.251802},
	url          = {https://link.aps.org/doi/10.1103/PhysRevLett.127.251802},
	urldate      = {2024-02-26}
}

@article{marshall_directional_2021,
	title        = {Directional detection of dark matter with diamond},
	author       = {Marshall, Mason C. and Turner, Matthew J. and Ku, Mark J. H. and Phillips, David F. and Walsworth, Ronald L.},
	year         = 2021,
	month        = mar,
	journal      = {Quantum Science and Technology},
	volume       = 6,
	number       = 2,
	pages        = {024011},
	doi          = {10.1088/2058-9565/abe5ed},
	urldate      = {2024-01-23}
}

@article{kehayias_crystalstress_2019,
	title        = {Imaging crystal stress in diamond using ensembles of nitrogen-vacancy centers},
	author       = {Kehayias, P. and Turner, M. J. and Trubko, R. and Schloss, J. M. and Hart, C. A. and Wesson, M. and Glenn, D. R. and Walsworth, R. L.},
	year         = 2019,
	month        = {Nov},
	journal      = {Physical Review B},
	publisher    = {American Physical Society},
	volume       = 100,
	pages        = 174103,
	doi          = {10.1103/PhysRevB.100.174103},
	url          = {https://link.aps.org/doi/10.1103/PhysRevB.100.174103},
	issue        = 17,
	numpages     = 8
}

@article{ebadi_directional_2022,
	title        = {Directional detection of dark matter using solid-state quantum sensing},
	author       = {Ebadi, Reza and Marshall, Mason C. and Phillips, David F. and Cremer, Johannes and Zhou, Tao and Titze, Michael and Kehayias, Pauli and Saleh Ziabari, Maziar and Delegan, Nazar and Rajendran, Surjeet and Sushkov, Alexander O. and Heremans, F. Joseph and Bielejec, Edward S. and Holt, Martin V. and Walsworth, Ronald L.},
	year         = 2022,
	month        = nov,
	journal      = {AVS Quantum Science},
	volume       = 4,
	number       = 4,
	pages        = {044701},
	doi          = {10.1116/5.0117301},
	urldate      = {2024-01-23}
}

@article{baum_mineral_2023,
	title        = {Mineral detection of neutrinos and dark matter. {A} whitepaper},
	author       = {Baum, Sebastian and Stengel, Patrick and Abe, Natsue and Acevedo, Javier F. and Araujo, Gabriela R. and Asahara, Yoshihiro and Avignone, Frank and Balogh, Levente and Baudis, Laura and Boukhtouchen, Yilda and Bramante, Joseph and Breur, Pieter Alexander and Caccianiga, Lorenzo and Capozzi, Francesco and Collar, Juan I. and Ebadi, Reza and Edwards, Thomas and Eitel, Klaus and Elykov, Alexey and Ewing, Rodney C. and Freese, Katherine and Fung, Audrey and Galelli, Claudio and Glasmacher, Ulrich A. and Gleason, Arianna and Hasebe, Noriko and Hirose, Shigenobu and Horiuchi, Shunsaku and Hoshino, Yasushi and Huber, Patrick and Ido, Yuki and Igami, Yohei and Ishikawa, Norito and Itow, Yoshitaka and Kamiyama, Takashi and Kato, Takenori and Kavanagh, Bradley J. and Kawamura, Yoji and Kazama, Shingo and Kenney, Christopher J. and Kilminster, Ben and Kouketsu, Yui and Kozaka, Yukiko and Kurinsky, Noah A. and Leybourne, Matthew and Lucas, Thalles and McDonough, William F. and Marshall, Mason C. and Mateos, Jose Maria and Mathur, Anubhav and Michibayashi, Katsuyoshi and Mkhonto, Sharlotte and Murase, Kohta and Naka, Tatsuhiro and Oguni, Kenji and Rajendran, Surjeet and Sakane, Hitoshi and Sala, Paola and Scholberg, Kate and Semenec, Ingrida and Shiraishi, Takuya and Spitz, Joshua and Sun, Kai and Suzuki, Katsuhiko and Tanin, Erwin H. and Vincent, Aaron and Vladimirov, Nikita and Walsworth, Ronald L. and Watanabe, Hiroko},
	year         = 2023,
	month        = aug,
	journal      = {Physics of the Dark Universe},
	volume       = 41,
	pages        = 101245,
	doi          = {10.1016/j.dark.2023.101245},
	urldate      = {2024-01-23}
}

@article{marshall_high-precision_2022,
	title        = {High-{Precision} {Mapping} of {Diamond} {Crystal} {Strain} {Using} {Quantum} {Interferometry}},
	author       = {Marshall, Mason C. and Ebadi, Reza and Hart, Connor and Turner, Matthew J. and Ku, Mark J.H. and Phillips, David F. and Walsworth, Ronald L.},
	year         = 2022,
	journal      = {Physical Review Applied},
	volume       = 17,
	number       = 2,
	pages        = {024041},
	doi          = {10.1103/PhysRevApplied.17.024041}
}

@article{kurinsky_diamond_2019,
	title        = {Diamond detectors for direct detection of sub-{GeV} dark matter},
	author       = {Kurinsky, Noah and Yu, To Chin and Hochberg, Yonit and Cabrera, Blas},
	year         = 2019,
	month        = jun,
	journal      = {Physical Review D},
	volume       = 99,
	number       = 12,
	pages        = 123005,
	doi          = {10.1103/PhysRevD.99.123005}
}

@article{SangtawesinDeLeon2019surfaceprep,
	title        = {Origins of Diamond Surface Noise Probed by Correlating Single-Spin Measurements with Surface Spectroscopy},
	author       = {Sangtawesin, Sorawis and Dwyer, Bo L. and Srinivasan, Srikanth and Allred, James J. and Rodgers, Lila V. H. and De Greve, Kristiaan and Stacey, Alastair and Dontschuk, Nikolai and O'Donnell, Kane M. and Hu, Di and Evans, D. Andrew and Jaye, Cherno and Fischer, Daniel A. and Markham, Matthew L. and Twitchen, Daniel J. and Park, Hongkun and Lukin, Mikhail D. and de Leon, Nathalie P.},
	year         = 2019,
	month        = {Sep},
	journal      = {Physical Review X},
	publisher    = {American Physical Society},
	volume       = 9,
	pages        = {031052},
	doi          = {10.1103/PhysRevX.9.031052},
	url          = {https://link.aps.org/doi/10.1103/PhysRevX.9.031052},
	issue        = 3,
	numpages     = 17
}

@article{umemoto2023scintillation,
	title        = {Basic characteristics of synthetic-diamond scintillator},
	author       = {Atsuhiro Umemoto and Takashi Iida and Masao Yoshino and Akira Yoshikawa and Shintaro Nomura},
	year         = 2023,
	journal      = {Nuclear Instruments and Methods in Physics Research Section A: Accelerators, Spectrometers, Detectors and Associated Equipment},
	volume       = 1057,
	pages        = 168789,
	doi          = {https://doi.org/10.1016/j.nima.2023.168789},
	issn         = {0168-9002},
	url          = {https://www.sciencedirect.com/science/article/pii/S0168900223007805}
}

@article{Drukier_DMminerals_2019,
	title        = {Paleo-detectors: Searching for dark matter with ancient minerals},
	author       = {Drukier, Andrzej K. and Baum, Sebastian and Freese, Katherine and G\'orski, Maciej and Stengel, Patrick},
	year         = 2019,
	month        = {Feb},
	journal      = {Physical Review D},
	publisher    = {American Physical Society},
	volume       = 99,
	pages        = {043014},
	doi          = {10.1103/PhysRevD.99.043014},
	url          = {https://link.aps.org/doi/10.1103/PhysRevD.99.043014},
	issue        = 4,
	numpages     = 23
}

@article{holtNanoscaleHardXRay2013,
	title        = {Nanoscale {Hard} {X}-{Ray} {Microscopy} {Methods} for {Materials} {Studies}},
	author       = {Holt, Martin and Harder, Ross and Winarski, Robert and Rose, Volker},
	year         = 2013,
	journal      = {Annual Review of Materials Research},
	volume       = 43,
	number       = 1,
	pages        = {183--211},
	doi          = {10.1146/annurev-matsci-071312-121654},
	url          = {https://doi.org/10.1146/annurev-matsci-071312-121654},
	urldate      = {2024-02-27},
	note         = {\_eprint: https://doi.org/10.1146/annurev-matsci-071312-121654}
}

@article{levine_principles_2019,
	title        = {Principles and techniques of the quantum diamond microscope},
	author       = {Levine, Edlyn V. and Turner, Matthew J. and Kehayias, Pauli and Hart, Connor A. and Langellier, Nicholas and Trubko, Raisa and Glenn, David R. and Fu, Roger R. and Walsworth, Ronald L.},
	year         = 2019,
	month        = nov,
	journal      = {Nanophotonics},
	volume       = 8,
	number       = 11,
	pages        = {1945--1973},
	doi          = {10.1515/nanoph-2019-0209},
	issn         = {2192-8614},
	url          = {https://www.degruyter.com/document/doi/10.1515/nanoph-2019-0209/html?lang=en},
	urldate      = {2023-09-16},
	copyright    = {De Gruyter expressly reserves the right to use all content for commercial text and data mining within the meaning of Section 44b of the German Copyright Act.},
	language     = {en}
}

@article{amawiThreedimensionalMagneticResonance2024,
	title        = {Three-dimensional magnetic resonance tomography with sub-10 nanometer resolution},
	author       = {Amawi, Mohammad T. and Trelin, Andrii and Huang, You and Weinbrenner, Paul and Poggiali, Francesco and Leibold, Joachim and Schalk, Martin and Reinhard, Friedemann},
	year         = 2024,
	month        = jan,
	journal      = {npj Quantum Information},
	volume       = 10,
	number       = 1,
	pages        = 16,
	doi          = {10.1038/s41534-024-00809-w},
	issn         = {2056-6387},
	url          = {https://www.nature.com/articles/s41534-024-00809-w},
	urldate      = {2025-10-31},
	copyright    = {2024 The Author(s)},
	language     = {en}
}

@article{Alsid2019nvcreation,
	title        = {Photoluminescence Decomposition Analysis: A Technique to Characterize {$\mathrm{N}$-$V$} Creation in Diamond},
	author       = {Alsid, Scott T. and Barry, John F. and Pham, Linh M. and Schloss, Jennifer M. and O'Keeffe, Michael F. and Cappellaro, Paola and Braje, Danielle A.},
	year         = 2019,
	month        = {Oct},
	journal      = {Physical Review Applied},
	publisher    = {American Physical Society},
	volume       = 12,
	pages        = {044003},
	doi          = {10.1103/PhysRevApplied.12.044003},
	url          = {https://link.aps.org/doi/10.1103/PhysRevApplied.12.044003},
	issue        = 4,
	numpages     = 20
}

@article{beyerElectronDensityThermal2023,
	title        = {Electron density and thermal motion of diamond at elevated temperatures},
	author       = {Beyer, J. and Grønbech, T. B. E. and Zhang, J. and Kato, K. and Brummerstedt Iversen, B.},
	year         = 2023,
	month        = jan,
	journal      = {Acta Crystallographica Section A: Foundations and Advances},
	publisher    = {International Union of Crystallography},
	volume       = 79,
	number       = 1,
	pages        = {41--50},
	doi          = {10.1107/S2053273322010154},
	issn         = {2053-2733},
	url          = {https://journals.iucr.org/a/issues/2023/01/00/pl5020/},
	urldate      = {2026-05-10},
	language     = {en}
}

@article{marshall_xray_2021,
	title        = {Scanning {X}-{Ray} {Diffraction} {Microscopy} for {Diamond} {Quantum} {Sensing}},
	author       = {Marshall, Mason C. and Phillips, David F. and Turner, Matthew J. and Ku, Mark J. H. and Zhou, Tao and Delegan, Nazar and Heremans, F. Joseph and Holt, Martin V. and Walsworth, Ronald L.},
	year         = 2021,
	month        = nov,
	journal      = {Physical Review Applied},
	volume       = 16,
	number       = 5,
	pages        = {054032},
	doi          = {10.1103/PhysRevApplied.16.054032},
	url          = {https://link.aps.org/doi/10.1103/PhysRevApplied.16.054032},
	urldate      = {2023-09-16}
}

@article{Bauch2020,
	title        = {Decoherence of ensembles of nitrogen-vacancy centers in diamond},
	author       = {Bauch, Erik and Singh, Swati and Lee, Junghyun and Hart, Connor A. and Schloss, Jennifer M. and Turner, Matthew J. and Barry, John F. and Pham, Linh M. and Bar-Gill, Nir and Yelin, Susanne F. and Walsworth, Ronald L.},
	year         = 2020,
	month        = {Oct},
	journal      = {Physical Review B},
	publisher    = {American Physical Society},
	volume       = 102,
	pages        = 134210,
	doi          = {10.1103/PhysRevB.102.134210},
	url          = {https://link.aps.org/doi/10.1103/PhysRevB.102.134210},
	issue        = 13,
	numpages     = 9
}

@article{XENONsolarneutrino2024,
	title        = {First Indication of Solar $^{8}\mathrm{B}$ Neutrinos via Coherent Elastic Neutrino-Nucleus Scattering with {XENONnT}},
	author       = {{XENON Collaboration}},
	year         = 2024,
	month        = {Nov},
	journal      = {Phys. Rev. Lett.},
	publisher    = {American Physical Society},
	volume       = 133,
	pages        = 191002,
	doi          = {10.1103/PhysRevLett.133.191002},
	url          = {https://link.aps.org/doi/10.1103/PhysRevLett.133.191002},
	issue        = 19,
	numpages     = 11
}

@article{nordlund2018,
  author  = {Nordlund, K. and Zinkle, S. J. and Sand, A. E. and Granberg, F. and
             Averback, R. S. and Stoller, R. E. and Suzudo, T. and Malerba, L. and
             Banhart, F. and Weber, W. J. and Willaime, F. and Dudarev, S. L. and
             Simeone, D.},
  title   = {Primary radiation damage: A review of current understanding and models},
  journal = {Journal of Nuclear Materials},
  volume  = {512},
  pages   = {450--479},
  year    = {2018},
  doi     = {10.1016/j.jnucmat.2018.10.027},
}

@article{manson2018,
  title   = {{NV$^-$--N$^+$} pair centre in {1b} diamond},
  author  = {Manson, N. B. and Hedges, M. and Barson, M. S. J. and Ahlefeldt, R. and Doherty, M. W. and Abe, H. and Ohshima, T. and Sellars, M. J.},
  year    = 2018,
  journal = {New Journal of Physics},
  volume  = 20, pages = {113037},
  doi     = {10.1088/1367-2630/aaec58}
}

\end{document}